\documentclass[lettersize,journal]{IEEEtran}
\usepackage{amsmath}
\usepackage{amsthm}
\usepackage{amssymb}
\usepackage{graphicx}
\usepackage{algorithm}
\usepackage{algorithmic}
\usepackage{color}
\usepackage{cite}
\usepackage[caption=false]{subfig}
\newtheorem{theorem}{Theorem}
\newtheorem{lemma}{Lemma}

\ifCLASSINFOpdf
\else
\fi

\begin{document}
\title{Doppler Shift Keying Modulation for Uplink Multiple Access over Doubly-Dispersive Channels}
	
\author{
	Xuehan~Wang,~\IEEEmembership{Graduate~Student~Member,~IEEE},
	Jintao~Wang,~\IEEEmembership{Fellow,~IEEE},
	Hai~Lin,~\IEEEmembership{Senior~Member,~IEEE},
	Jinhong~Yuan,~\IEEEmembership{Fellow,~IEEE},
	Xu~Shi,~\IEEEmembership{Member,~IEEE},
	Hengyu~Zhang,~\IEEEmembership{Graduate~Student~Member,~IEEE},~and
	Jian~Song,~\IEEEmembership{Fellow,~IEEE}
	\thanks{This work was supported in part by Beijing Natural Science Foundation under Grant 4254069, in part by the Project funded by China Postdoctoral Science Foundation under Grant 2024M761671, and also in part by the Japan Society for the  Promotion of Science (JSPS) Grants-in-Aid for Scientific Research (KAKENHI) under Grant 22H01491. \textit{(Corresponding author: Jintao Wang.)}\par 
	Xuehan Wang, Jintao Wang, Xu Shi and Hengyu Zhang are with the Department of Electronic Engineering, Tsinghua University, Beijing 100084, China (e-mail: wang-xh21@mails.tsinghua.edu.cn; wangjintao@tsinghua.edu.cn; shi-x@tsinghua.edu.cn; zhanghen23@mails.tsinghua.edu.cn).\par 
	Hai Lin is with the Department of Electrical and Electronic Systems Engineering, Graduate School of Engineering, Osaka Metropolitan University, Sakai, Osaka 599-8531, Japan (e-mail: hai.lin@ieee.org).\par
	Jinhong Yuan is with the School of Electrical Engineering and Telecommunications, University of New South Wales, Sydney, NSW 2052, Australia (e-mail: j.yuan@unsw.edu.au).\par 
	Jian Song is with the Tsinghua Shenzhen International Graduate School, Shenzhen 518055, China  (e-mail: jsong@tsinghua.edu.cn).
	}
}
	
\maketitle
\begin{abstract} 
The delay-Doppler (DD) domain modulation has been regarded as one of the most competitive candidates to support wireless communications for emerging high-mobility applications in the sixth-generation mobile networks. Unfortunately, most of the existing designs for DD domain modulation suffer from high peak-to-average power ratio (PAPR) and unbearable detection complexity under uplink transmission since large time duration and bandwidth are required to guarantee high DD resolutions. To address these issues, the Doppler shift keying (DSK) modulation based on the orthogonal delay Doppler division multiplexing modulator is proposed in this paper, where the input-output characterization in the DD domain is fully exploited. The principle of the DSK transceiver is first established with the one-hot mapper and low-complexity iterative successive interference cancellation-maximum ratio combining detector for point-to-point scenarios. The proposed scheme is then generalized to the zero auto-correlation sequence-based implementation, which benefits the extension of multi-user (MU) uplink DSK frameworks. For uplink DSK transmission, Zadoff-Chu (ZC) sequences are adopted as the basis sequences. We optimize the assignment of ZC roots to different user equipments (UEs) by minimizing the maximum inter-user interference. This optimization process, which analyzes the root allocation, directly assigns a specific ZC sequence to each UE. The PAPR and bit error rate performance of the proposed DSK modulation with the low-complexity detector is finally verified by extensive simulation results under doubly-dispersive channels, which demonstrates the superiority of DSK modulation especially for uplink multiple access over doubly dispersive channels.
\end{abstract}
	
\IEEEpeerreviewmaketitle
\section{Introduction}
\label{sec:intro}
The sixth-generation (6G) mobile networks are envisioned to support ultra-reliable data transmission under high-mobility scenarios \cite{OTFS_CE_onlineBL_mobility,ODDM_recent,ref_OTFS_magazine_wei_mobility,ref_mmWave_mobility}, which serves as one of the most essential technologies in intelligent transportation systems \cite{DL_ITS,DDMA_ITS} and non-terrestrial networks \cite{ref_NTN1,ref_NTN2,ref_NTN3}. Regrettably, significant reliability loss appears in existing modulation techniques deployed in the fifth-generation new radio (5G-NR) systems, e.g., orthogonal frequency division multiplexing (OFDM). The reason behind it is the doubly-dispersive channel brought by the multipath environment and high mobility, which leads to severe selective fading in the time-frequency domain and destroys the orthogonality among subcarriers.\par
Recently, some innovations of modulation schemes resilient to doubly-dispersive channels have been proposed, e.g., the delay-Doppler (DD) domain modulation in \cite{OTFS_SU_MPA,ref_ODDM,ODDM_tutorial,Yuan_CHina}, orthogonal time sequency multiplexing (OTSM) modulation in \cite{ref_OTSM} and affine frequency division multiplexing (AFDM) modulation in \cite{ref_AFDM}. Among various frameworks, the DD domain modulation schemes such as the orthogonal time frequency space (OTFS) modulation in \cite{OTFS_SU_MPA} and the orthogonal delay Doppler division multiplexing (ODDM) modulation in \cite{ref_ODDM,ODDM_tutorial} are more preferred due to the better coupling with the sparse DD domain channels, which offers significant convenience for channel estimation especially with massive multiple-input-multiple-output (MIMO) involved \cite{OTFS_MIMO,MIMO_ref}. On the other hand, by multiplexing data symbols in the DD domain, full time-frequency diversity can be acquired for each component, which provides insight into enhancing the reliability of wireless communication networks further \cite{OTFS_performance_P_TWC}. As a result, DD domain modulation has been widely regarded as a promising waveform candidate to complement the OFDM systems over doubly-dispersive channels \cite{compare_OTFS}. \par
To exploit the potential of DD domain modulation systems, the equivalent sampled delay-Doppler (ESDD) channel should be sparse \cite{ODDM_tutorial,OTFS_wideband_mine,ODDM_precoding}, which indicates that the bandwidth and time duration are required to be sufficiently large to guarantee high DD resolutions. Therefore, the computational load of DD domain detectors is extremely high if directly deploying traditional schemes, e.g., the linear minimum mean squared error (LMMSE) approach. To address this issue, the message passing-based algorithm (MPA) was deployed in OTFS systems in \cite{OTFS_SU_MPA,Yuan_commL}. Nevertheless, the complexity of MPA increases significantly with modulation order \cite{OTFS_newmodel}, which indicates the superiority of iterative linear detectors in \cite{OTFS_newmodel,OTFS_MRCdetect_TVT,OTFS_MA_Tcom_LMMSE_ZP,Lowcomdet_OTFS_ZP,OTFS_linear_TWC}. By simplifying the input-output relation with a zero padding (ZP) and grouping the received components into vectors according to their multicarrier symbol indices, a low-complexity iterative maximal ratio combining (MRC) detector was proposed in \cite{OTFS_MRCdetect_TVT}. The basic idea was developed further in \cite{Lowcomdet_OTFS_ZP} by adopting the turbo successive interference cancellation (SIC)-LMMSE detector with elaborate consideration of the channel decoder. Meanwhile, since deep learning (DL) has been demonstrated to be efficient in the optimization of wireless communications in \cite{DL_wireless,dl_reception_rui,dl_feedback_xudong,AI_communication}, DL-based design has also been fully exploited for DD domain detectors in \cite{ODDM_MIMO_DL,OTFS_DL,ODDM_cheng_simu_tcom}, which benefits the trade-off between the implementation complexity and detection reliability significantly. However, almost all of the aforementioned schemes cannot achieve satisfying performance for uplink transmission due to the more complicated multi-user interference (MUI) pattern in the DD domain \cite{OTFS_MA_Tcom_LMMSE_ZP}. As a result, more elaborate transmission designs are required if the reliability superiority enabled by DD domain transmission is expected to be exploited in uplink multiple access (MA) systems. \par
On the other hand, the high peak-to-average power ratio (PAPR) is also widely regarded as one of the unsettled barriers that prevent the practical realization of DD domain modulation \cite{OTFS_CE_embedded,OTFS_PAPR1,DCT_OTFS_PAPR_CL} especially in uplink transmission. Fortunately, the emerging techniques of shift keying (SK) and index modulation (IM) can mitigate the PAPR significantly \cite{OFDM_FSK,SSK_our,OTFS_STSK_Tcom}, which were investigated in the DD domain in \cite{OTFS_IM1,OTFS_IM2,OTFSMA_IM,DDM_IM_Tcom}. However, the input-output relation in the DD domain was not fully considered in these schemes, which resulted in the failure of existing iterative linear detectors since the data components cannot be grouped according to their multicarrier symbol indices like \cite{OTFS_MRCdetect_TVT,OTFS_MA_Tcom_LMMSE_ZP,Lowcomdet_OTFS_ZP,OTFS_linear_TWC,ref_OTFS_latest}. Therefore, the reduction of PAPR without appending much higher detection complexity remains a core problem for the DD domain frameworks. \par  
Motivated by the low PAPR of SK approaches \cite{OFDM_FSK} and the excellent trade-off between the complexity and reliability of iterative SIC-MRC detectors \cite{OTFS_MRCdetect_TVT}, the Doppler shift keying (DSK) modulation based on the orthogonal delay Doppler division multiplexing modulator is proposed in this paper to address above challenges by fully exploiting the input-output characterization in the DD domain. Specifically, subcarrier indices with respect to the Doppler resolution are adopted to load information bits by the SK method within each multicarrier symbol, which enables the low-complexity iterative detector and extremely low PAPR. The discussion can be extended to the multi-user (MU) uplink systems by adopting the optimized zero auto-correlation (ZAC) sequences, whose performance superiority is demonstrated by simulation results. The major contributions of this work are summarized as follows:
\begin{itemize}
	\item \textbf{Framework of DSK modulation:} Considering the input-output characterization in the DD domain, data components are grouped into vectors according to multicarrier symbol indices since the input-output relation within each multicarrier symbol can be treated as circular convolution thanks to the property of ODDM modulator/demodulator. The information bits are mapped into a one-hot constellation vector for each multicarrier symbol according to the corresponding subcarrier index. At the receiver, the matching pursuit (MP)-based detector and iterative SIC-MRC detector are developed with low complexity, which completes the basic framework of DSK modulation systems. 
	\item \textbf{Basis sequences-based implementation:} The concepts of the DSK modulation is then generalized to the basis sequence-based one, where the ZAC sequence is circularly shifted according to the corresponding index utilized to bear information bits. The proposed scheme is then extended to MU uplink scenarios by allocating different user equipments (UEs) with different ZAC sequences, where the MU detector is also proposed. It is worth pointing out that the proposed iterative linear detection scheme provides excellent performance in a parallel manner, which is significantly different from existing designs in \cite{Lowcomdet_OTFS_ZP,OTFS_MRCdetect_TVT} and reduces the computation complexity significantly.
	\item \textbf{Optimization of MU uplink basis sequences:} Considering the outstanding property of Zadoff-Chu (ZC) sequences \cite{ref_ZC_chu}, ZC sequences with different roots are adopted as basis sequences for different UEs in this paper. The closed-form optimization of the sequence roots is derived from the analysis of the cross-correlation among different UEs, which determines the level of MUI by investigating the received signal at the base station (BS) side. The excellent PAPR and bit error rate (BER) performance of the proposed DSK modulation is finally demonstrated by executing numerical experiments under doubly-dispersive channels.
\end{itemize}\par
Since the level of MUI is governed by the cross-correlation among MU basis sequences, the proposed root optimization effectively suppresses MUI. Consequently, each UE can approximately preserve its individual transmission rate without requiring resource partitioning. As a result, the overall system throughput increases nearly linearly with the number of UEs, in contrast to conventional ODDM systems, where resource allocation leads to reduced throughput for each UE. Meanwhile, the mitigated MUI also helps reduce the detection latency by parallel processing.\par  
The rest of this paper is organized as follows. The system model is first briefly reviewed in Section \ref{sec:system}. The transceiver design of the proposed DSK modulation is illustrated in Section \ref{sec:P2P} for point-to-point systems. The discussion is then extended to uplink DSK transmission in Section \ref{sec:uplink}. The performance evaluation is provided in Section \ref{sec:simu}. Finally, conclusions are briefly drawn in Section \ref{sec:conclusion}.\par
\textit{Notations}: $\mathcal{A}$ is a set, $\mathbf{A}$ is a matrix, $\mathbf{a}$ is a column vector, $a$ is a scalar. $\mathbf{a}_{i}$ represents the $i$-th column of $\mathbf{A}$ while $\mathbf{A}(i,j)$ stands for the $(i,j)$ component of $\mathbf{A}$. $\mathbf{A}^H$ denotes its conjugate transposition. $(\cdot)^{*}$ stands for the conjugate operation, while $(\cdot)_{N}$ represents the modulus operation with respect to $N$. $\text{gcd}(a,b)$ denotes the greatest common divisor of the elements of $a$ and $b$. Finally, $\mathbf{I}_{N}$ and $\mathbf{F}_{N}$ stand for the $N\times N$ identity matrix and normalized $N$-point discrete Fourier Transform (DFT) matrix, respectively.
\section{System Model}
\label{sec:system}
In this section, the high-mobility channel model and ODDM modulation/demodulation are briefly reviewed, which serve as the basis of the proposed DSK framework. Let $\tau_{r}$ and $\nu_{r}$ respectively denote the delay resolution and Doppler resolution, which requires the effective bandwidth $B_{s}$ and time duration $T_{s}$ of the baseband transmit signal $s(t)$ to satisfy $B_{s}>\frac{1}{\tau_{r}}$ and $T_{s}>\frac{1}{\nu_{r}}$, respectively. For wireless communications with practical pulse-shaping, $\tau_{r}$ is equal to the Nyquist sampling period. $N$ and $M$ represent the number of ODDM subcarriers and multicarrier symbols, respectively, which leads to the total ODDM frame duration of $NM\tau_{r}$ since there will be $NM$ time domain samples. Therefore, the aforementioned DD resolutions can be acquired by $\nu_{r}=\frac{1}{NM\tau_{r}}$. Finally, the data components to be modulated and the demodulated components at the $k$-th subcarrier within the $l$-th multicarrier symbol are respectively denoted as $\left\{x[k,l]|k=0,\cdots, N-1, l=0,\cdots, M-1\right\}$ and $\left\{y[k,l]|k=0,\cdots, N-1, l=0,\cdots, M-1\right\}$.\par 

\subsection{Doubly-Dispersive Channel Model}
\label{subsec:channelModel}
In this subsection, the multipath channel model with high mobility is provided, which reveals the doubly-dispersive property. The impact of noise is disregarded for ease of illustration. After the up-conversion, the passband signal $\tilde{s}(t)=\Re\left\{s(t)e^{j2\pi f_{c}t}\right\}$ is sent from the transmitter to the receiver via $\tilde{P}$ incident paths, where $f_{c}$ denotes the carrier frequency while $s(t)$ represents the continuous-time baseband waveform. According to \cite{ref_LTVchannel_book,OTFS_wideband_mine,OTFS_DSE_GC_mine}, the received passband signal can be derived as 
\begin{equation}
	\tilde{r}(t)=\Re\left\{\sum_{p=1}^{\tilde{P}}\tilde{h}_{p}^{\prime}s(t-(\tilde{\tau}_{p}-\frac{\tilde{v}_{p}}{\text{c}}t))e^{j2\pi f_{c}\left(t-(\tilde{\tau}_{p}-\frac{\tilde{v}_{p}}{\text{c}}t)\right)}\right\},
	\label{passband_io_continue}
\end{equation}
where $\tilde{v}_{p}$, $\tilde{\tau}_{p}$ and $\tilde{h}_{p}^{\prime}$, respectively, stand for the velocity with which the path length is decreasing, propagation delay, and attenuation associated with the $p$-th path. $\text{c}$ represents the speed of light. Let $\tilde{\nu}_{p}=\frac{\tilde{v}_{p}}{\text{c}}f_{c}$ denote the Doppler shift of the $p$-th path corresponding to the carrier frequency. After removing the carrier, the baseband received signal can be derived as
\begin{equation}
	r(t)=\sum_{p=1}^{\tilde{P}}\tilde{h}_{p}e^{j2\pi\tilde{\nu}_{p}t}s\left(t-(\tilde{\tau}_{p}-\frac{\tilde{v}_{p}}{\text{c}}t)\right),
	\label{banseband_continue_dse}
\end{equation}  
where we have $\tilde{h}_{p}=\tilde{h}_{p}^{\prime}e^{-j2\pi f_{c}\tilde{\tau}_{p}}$. Throughout this paper, the narrowband assumption\footnote{Please note that this condition does not indicate that the Doppler shift or time delay is banded. Instead, it is a necessary condition to derive typical doubly-dispersive channels since the relative mobility should be characterized as the Doppler scaling effect if the narrowband assumption is not satisfied. In DD domain modulation systems, this condition can also be transferred into $\left|\frac{\tilde{v}_{p}}{c}NM\right|\ll1$, and this is also a necessary condition to utilize DD domain modulation schemes.} is satisfied as \cite{ODDM_tutorial}, i.e., the time-variant delay $\frac{\tilde{v}_{p}}{\text{c}}t$ is far less than the delay resolution $\tau_{r}$ during the whole duration of $s(t)$. Therefore, \eqref{banseband_continue_dse} can be rewritten as 
\begin{equation}
	r(t)=\sum_{p=1}^{\tilde{P}}\tilde{h}_{p}e^{j2\pi\tilde{\nu}_{p}t}s(t-\tilde{\tau}_{p}).
    \label{banseband_continue}
\end{equation}
Meanwhile, the band-limited filtering and sampling lead to an equivalent \textit{on-grid} channel as indicated in \cite{ref_ODDM,ODDM_tutorial,ref_channel_onGrid,ref_station_time,ODDM_ESDD_ref_jun_TCOM}, namely the ESDD channel model. The ESDD model corresponding to \eqref{banseband_continue} can be derived as
\begin{equation}
	r(t)=\int\int h(\tau,\nu)s(t-\tau)e^{j2\pi\nu t}d\tau d\nu.
	\label{DD_general}
\end{equation}
$h(\tau,\nu)$ in \eqref{DD_general} is the observed ESDD\footnote{Please note that the ESDD channel does not mean that the physical channel reveals on-grid delay and Doppler shift. In fact, a general physical channel leads to an equivalent ESDD channel \cite{ODDM_ESDD_ref_jun_TCOM} with $P\geq\tilde{P}$, which is already sufficient to implement the receiver design in general communication systems \cite{ref_ODDM,ODDM_tutorial,ref_channel_onGrid,ref_station_time}. As a result, we focus on the basic principle of DSK framework under ESDD channels whose performance has been verified in \cite{ODDM_cheng_simu_tcom}. Taking the initial physical parameters into account can further promote the performance, which will certainly be embodied in our future work.} channel characterized as
\begin{equation}
	h(\tau,\nu)=\sum_{p=1}^{P}h_{p}\delta(\tau-\tau_{p})\delta({\nu-\nu_{p}}),
	\label{ESDD}
\end{equation}
where we have $P\geq\tilde{P}$, $\tau_{p}=l_{p}\tau_{r}$, $\nu_{p}=k_{p}\nu_{r}$, $l_{p}\in\mathbb{N}$ and $k_{p}\in\mathbb{Z}$, respectively. To ensure the sparsity of \eqref{ESDD}, delay and Doppler resolutions are required to be small enough, which means $s(t)$ occupies both sufficiently large bandwidth and long duration. However, since the aforementioned derivations are attained under the stationary and narrowband assumptions like \cite{ODDM_tutorial,ref_station_time}, the signal bandwidth and duration are bounded, which indicates the trade-off when determining signal parameters\footnote{It is suggested to refer to our prior work in \cite{OTFS_wideband_mine,OTFS_DSE_GC_mine} to obtain the analysis when the narrowband assumption is not satisfied.}. Let $l_{\text{max}}=\max_{p=1,\cdots, P}{l_{p}}$ and $k_{\text{max}}=\max_{p=1,\cdots, P}|k_{p}|$ denote the maximum sampled delay time and Doppler shift, respectively.\par 
\begin{figure*}
	\centering{\includegraphics[width=0.76\linewidth]{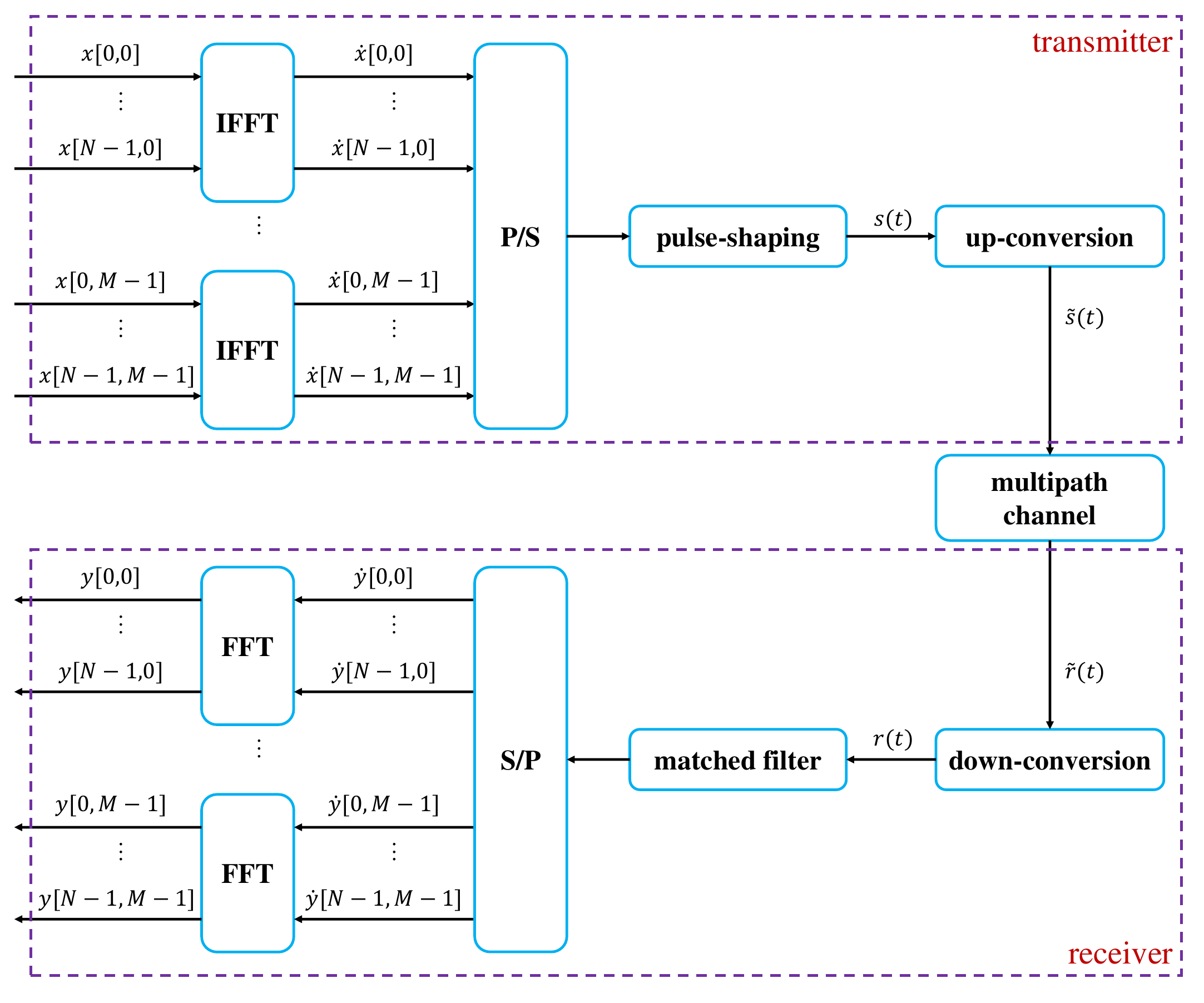}}
	\caption{Block diagram of the ODDM transmitter and receiver with the approximated low-complexity implementation.}
	\label{Fig0_ODDM}	
\end{figure*}
\subsection{DD Domain Signal Processing with ODDM}
In this subsection, the ODDM signal model and implementation are reviewed to serve as the basis. The ODDM modulation constructs the waveform by utilizing the conventional multicarrier processing \cite{ref_TFMC,multicarrier_survey,ODDM_tutorial,ref_ODDM_primer}. The baseband waveform can be generated as
\begin{equation}
	\label{ODDM_initial_mod}
	s(t)=\sum_{l=0}^{M-1}\sum_{k=0}^{N-1}x[k,l]g(t-l\tau_{r})e^{j2\pi k\nu_{r}(t-l\tau_{r})}.
\end{equation}
The modulation parameters are chosen such that the multicarrier symbol spacing $\tau_{r}$ and the subcarrier spacing $\nu_{r}$ correspond to DD resolutions, respectively. This design ensures that the synthesized multicarrier waveform is inherently matched to the DD domain channel representation. From the basic principle of ODDM modulation \cite{multicarrier_survey,ODDM_tutorial}, the prototype filter $g(t)$ in \eqref{ODDM_initial_mod} should guarantee the sufficient orthogonality with respect to DD resolutions as
\begin{equation}
	\label{DDOP_orthogonality}
	\begin{aligned}
		A_{g,g}(l\tau_{r},k\nu_{r})
		&=\int g^{*}(t-l\tau_{r})g(t)e^{-j2\pi k\nu_{r}(t-l\tau_{r})}dt\\
		&=\delta[l]\delta[k],~~|l|\leq M-1, |k|\leq N-1.
	\end{aligned} 
\end{equation}
Considering this property, the prototype filter $g(t)$ is also referred to as the DD domain orthogonal pulse (DDOP) in ODDM systems. Based on the analysis in \cite{ref_ODDM,ref_DDOP_ICC}, $g(t)$ can be derived by
\begin{equation}
	\label{DDOP_eq}
	g(t)=\frac{1}{\sqrt{N}}\sum_{n=0}^{N-1}a_{s}(t-nT),
\end{equation}
where $a_{s}(t)$ represents the truncated square-root Nyquist pulse for the symbol interval of $\tau_{r}$, and $T=M\tau_{r}$ denotes the interval between two pulses within the DDOP. The time duration of $a_{s}(t)$ is denoted as $2Q\tau_{r}$, where $2Q\ll{M}$ is required to avoid the extra complexity of cyclic extension as shown in \cite{ref_ODDM,ref_DDOP_ICC,ODDM_tutorial,ref_ODDM_primer}. Corresponding to the waveform generation at the transmitter in \eqref{ODDM_initial_mod}, the DD domain components can be obtained at the receiver side as
\begin{equation}
	\label{ODDM_initial_demod}
	y[k,l]=\int g^{*}(t-l\tau_{r})e^{-j2\pi k\nu_{r}(t-l\tau_{r})}r(t)dt.
\end{equation} 
However, the standard processing in \eqref{ODDM_initial_mod} and \eqref{ODDM_initial_demod} involves huge computational complexity due to the complicated form of DDOP. To reduce the implementation complexity, the baseband waveform $s(t)$ at the transmitter side can be generated by utilizing the inverse fast Fourier transform (IFFT) and sample-wise pulse-shaping with $a_{s}(t)$ as shown in Fig. \ref{Fig0_ODDM}. To be more specific, \eqref{ODDM_initial_mod} can be simplified as
\begin{equation}
	\label{ODDM_simplify_derivation}
	\begin{aligned}
		&s(t)=\sum_{l=0}^{M-1}\sum_{k=0}^{N-1}x[k,l]g(t-l\tau_{r})e^{j2\pi k\nu_{r}(t-l\tau_{r})}\\
		&=\frac{1}{\sqrt{N}}\sum_{l=0}^{M-1}\sum_{k=0}^{N-1}\sum_{n=0}^{N-1}x[k,l]a_{s}(t-l\tau_{r}-nT)e^{j2\pi\frac{k(t-l\tau_{r})}{NM\tau_{r}}}\\
		&\approx\frac{1}{\sqrt{N}}\sum_{l=0}^{M-1}\sum_{n=0}^{N-1}a_{s}(t-l\tau_{r}-nT)\sum_{k=0}^{N-1}x[k,l]e^{j2\pi\frac{knT}{NM\tau_{r}}}\\
		&=\sum_{l=0}^{M-1}\sum_{n=0}^{N-1}\dot{x}[n,l]a_{s}(t-l\tau_{r}-nT),
	\end{aligned}
\end{equation}
where the time domain samples can be obtained by employing the normalized $N$-point IFFT as
\begin{equation}
	\label{IFFT_ODDM_simplify}
	\dot{x}[n,l]=\frac{1}{\sqrt{N}}\sum_{k=0}^{N-1}x[k,l]e^{j2\pi\frac{kn}{N}}.
\end{equation}
Based on the illustrations above, the baseband waveform of the simplified ODDM system can be generated via two procedures. The time domain samples are first generated according to \eqref{IFFT_ODDM_simplify}. The transmit waveform can then be attained by the sample-wise pulse-shaping as
\begin{equation}
	\label{ODDM_pulseShaping}
	s(t)=\sum_{l=0}^{M-1}\sum_{n=0}^{N-1}\dot{x}[n,l]a_{s}(t-l\tau_{r}-nT).
\end{equation}\par
Accordingly, the matched filter is first executed at the receiver side to obtain the time domain samples as
\begin{equation}
	\label{MF_ODDM}
	\dot{y}[n,l]=\int a_{s}^{*}(t-l\tau_{r}-nT)r(t)dt.
\end{equation}\par
At last, the normalized $N$-point fast Fourier transform (FFT) is employed to attain DD domain components at the receiver side as
\begin{equation}
	\label{FFT_ODDM}
	y[k,l]=\frac{1}{\sqrt{N}}\sum_{n=0}^{N-1}\dot{y}[n,l]e^{-j2\pi\frac{nk}{N}}.
\end{equation}
\subsection{Input-Output Characterization in the DD Domain}
\label{subsec:DDIO}
Based on the DD domain signal processing from \eqref{IFFT_ODDM_simplify} to \eqref{FFT_ODDM} and the doubly-dispersive channel model illustrated in \eqref{DD_general} and \eqref{ESDD}, the input-output relation between $y[k,l]$ and $x[k,l]$ can be characterized as \cite{ODDM_tutorial,ref_ODDM}
\begin{equation}
	y[k,l]=\sum_{p=1}^{P}H_{p}[k,l]x[(k-k_{p})_{N},(l-l_{p})_{M}]+w[k,l],
	\label{DDIO_begin}
\end{equation}
where $w[k,l]\overset{\text{i.i.d.}}{\sim}\mathcal{CN}(0,\sigma^{2})$ is the additive white Gaussian noise samples. Based on the basic principle of ODDM and the multipath channel model in \eqref{ESDD}, $H_{p}[k,l]$ can be derived as
\begin{equation}
	H_{p}[k,l]=h_{p}e^{j2\pi\frac{k_{p}l}{NM}}\times
	\begin{cases}
		1,&l\geq l_{p}\\
		e^{-j2\pi\frac{k-k_{p}}{N}},&\text{elsewhere}
	\end{cases}.
\end{equation}
To simplify the system design, last $l_{\text{max}}$ multicarrier symbols are forced to zeros\footnote{ZP is utilized to simplify the analysis by guaranteeing the circular shift along the subcarriers as shown in \eqref{DDIO_final}. It can also be translated into no inter-sample-interference within each ODDM multicarrier symbol. By exploiting this property, ZAC sequences can be placed along ODDM subcarriers within each ODDM multicarrier symbol, which eases the interference mitigation and the corresponding detection design.} like \cite{OTFS_MRCdetect_TVT,Lowcomdet_OTFS_ZP,OTFS_MA_Tcom_LMMSE_ZP}, i.e., $x[k,l]=0$ holds for $l\geq M-l_{\text{max}}$. Therefore, \eqref{DDIO_begin} can be rewritten as
\begin{equation}
	y[k,l]=\sum_{p=1}^{P}h_{p}e^{j2\pi\frac{k_{p}l}{NM}}\mathbb{I}_{(l-l_{p})}x[(k-k_{p})_{N},l-l_{p}]+w[k,l],
	\label{DDIO_final}
\end{equation}
where $\mathbb{I}_{l}$ is the function that indicates whether the multicarrier symbol index $l$ is active. To be more specific, we have
\begin{equation}
	\mathbb{I}_{l}=\begin{cases}
		1,&0\leq l\leq M-l_{\text{max}}-1\\
		0,&\text{elsewhere}
	\end{cases}.
\end{equation}\par
The derivation in \eqref{DDIO_final} indicates that it is appropriate to group data components into vectors according to their multicarrier symbol indices, which means the term $e^{j2\pi\frac{k_{p}l}{NM}}$ can be absorbed to simplify the receiver design. It motivates the proposed DSK modulation design, which is illustrated in detail in the following sections. 
\begin{figure}
	\centering{\includegraphics[width=0.9\linewidth]{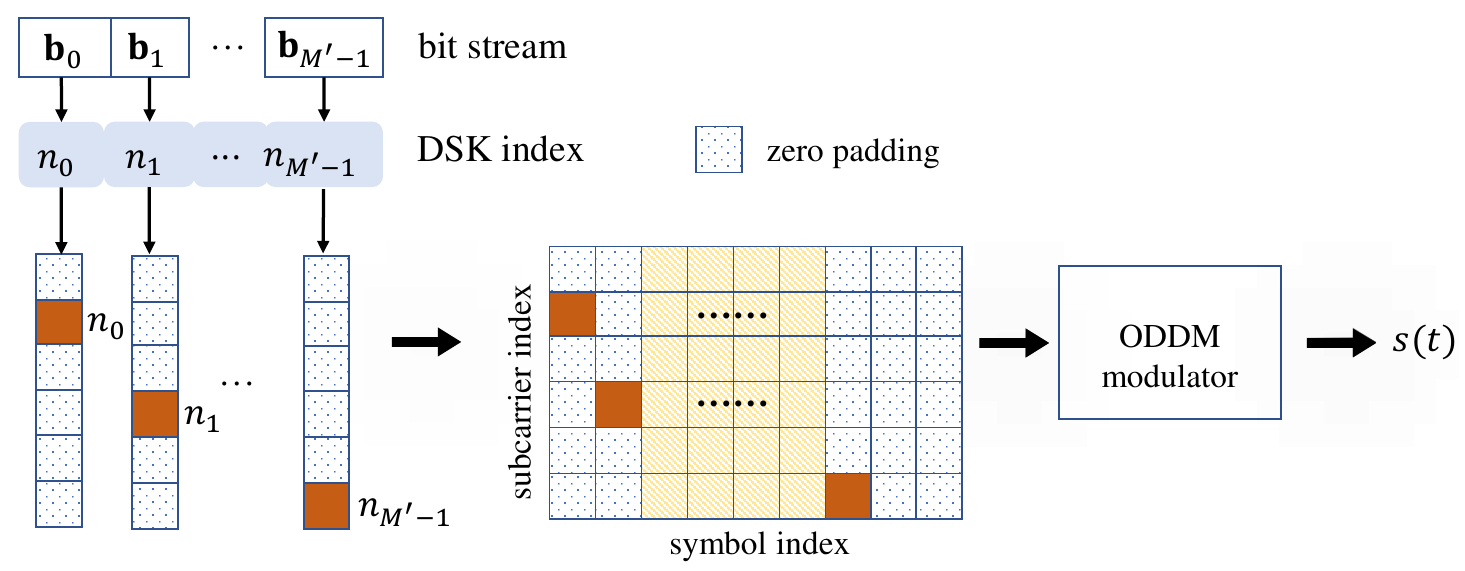}}
	\caption{Schematic of DSK transmitter with one-hot mapper.}
	\label{Fig_system_P2P}	
\end{figure}
\section{Point-to-Point DSK Transceivers}
\label{sec:P2P}
Inspired by the appreciable reliability promotion achieved by SSK modulation \cite{SSK_propose,OTFS_STSK_Tcom,SSK_our} in MIMO systems, the DSK framework is first proposed under point-to-point ODDM modulation systems considering the system model in ODDM systems presented in Section \ref{sec:system}. The discussion can be generalized to MU uplink scenarios as shown in Section \ref{sec:uplink}.
\subsection{DSK Transmitter}
\label{subsec:DSK_transmit}
Similar to the SSK modulation in MIMO systems, the subcarrier index with respect to the Doppler resolution is adopted to load data bits for each available multicarrier symbol index, whose transmitter model is illustrated in Fig. \ref{Fig_system_P2P}. The bit stream is first transferred into parallel bit sequences as $\mathbf{b}_{l}\in\{0,1\}^{N_{b}}$ for $\forall~\mathbb{I}_{l}=1$, where $N_{b}=\lfloor \log_{2}{N}\rfloor$ denotes the maximum number of bits for per multicarrier symbol index. After that, $\mathbf{b}_{l}$ is mapped into a constellation vector $\mathbf{x}_{l}=(x[0,l],x[1,l],\cdots,x[N-1,l])^{T}$ with the unit power constraint (i.e., $\mathbb{E}_{\mathbf{x}_{l}}\left(\mathbf{x}_{l}^{H}\mathbf{x}_{l}\right)=1$). Similar to \cite{SSK_propose,OTFS_STSK_Tcom}, the one-hot mapper is first selected to illustrate the basic principle of DSK modulation. Specifically, $\mathbf{b}_{l}$ is first encoded into the subcarrier index as $n_{l}\in[0,2^{N_{b}}-1]\subseteq[0,N-1]$. After that, $\mathbf{x}_{l}$ is generated as
\begin{equation}
	x[k,l]=\alpha_{l}\mathbb{I}_{l}\delta[k-n_{l}]=\begin{cases}
		\alpha_{l},&k=n_{l},~0\leq l\leq M^{\prime}-1\\
		0,&\text{elsewhere}
	\end{cases},
\end{equation}
where we have $|\alpha_{l}|=1$ and $M^{\prime}=M-l_{\text{max}}$. For ease of illustration, we set\footnote{Please note that although $\alpha_{l}$ does not contain information, it can be designed to optimize the transmission, which will be embodied in our future work.} $\alpha_{l}=1$ for $\forall~l$.\par
\begin{table}
	\caption{Example of the DSK mapper rule ($N=4$).}
	\centering
	\label{DSK_example_table}
	\renewcommand\arraystretch{1.3}
	\begin{tabular}{p{3em}|p{3em}|p{7em}|p{10em}}
		\hline
		$\mathbf{b}_{l}$ &
		$n_{l}$ &
		one-hot $\mathbf{x}_{l}$ &
		Sequence-based $\mathbf{x}_{l}$\\
		\hline
		$(0, 0)^{T}$ & 0 & $(1, 0, 0, 0)^{T}$ & $(z[0], z[1], z[2], z[3])^{T}$\\
		$(0, 1)^{T}$ & 1 & $(0, 1, 0, 0)^{T}$ & $(z[3], z[0], z[1], z[2])^{T}$\\
		$(1, 0)^{T}$ & 2 & $(0, 0, 1, 0)^{T}$ & $(z[2], z[3], z[0], z[1])^{T}$\\
		$(1, 1)^{T}$ & 3 & $(0, 0, 0, 1)^{T}$ & $(z[1], z[2], z[3], z[0])^{T}$\\
		\hline
	\end{tabular}
\end{table}
An example of the one-hot mapper-based DSK transmitter is shown in the first three columns of Table \ref{DSK_example_table}, where only one subcarrier index is activated per multicarrier symbol. After generating $\{x[k,l]\}$ according to the DSK mapper rule, the transmit waveform is attained by employing the ODDM modulator. As indicated in \cite{ref_station_time,ref_ODDM}, the waveform can be approximately treated as $N$-point inverse discrete Fourier transform (IDFT) along subcarrier indices and $a_{s}(t)$-based sample-wise filtering. Since the IDFT of the one-hot sequence has constant amplitude, the proposed DSK scheme enjoys better PAPR performance than traditional DD domain transmission approaches. \par
\subsection{Detector Design}
\label{subsec:DSK_receive}
At the receiver, $y[k,l]$ is first obtained by utilizing the ODDM demodulator. Considering the input-output characterization of \eqref{DDIO_final} and the DSK transmitter design, $y[k,l]$ can be derived as
\begin{equation}
	y[k,l]=\sum_{p=1}^{P}h_{p}e^{j2\pi\frac{k_{p}l}{NM}}\mathbb{I}_{l-l_{p}}\delta[(k-k_{p}-n_{l-l_{p}})_{N}]+w[k,l].
	\label{io_DSK_P2P}
\end{equation}\par 
If the multipath effect does not exist, $n_{l}$ can be recovered by directly employing the threshold-based detection. Considering the doubly-dispersive channel, a more elaborate detector design is required to guarantee reliability.\par 
At first, we provide an MP-based scheme by formulating the detector as a sparse signal recovery problem. The input-output relation in \eqref{io_DSK_P2P} can be vectorized as
\begin{equation}
	\mathbf{y}=\mathbf{Ax}+\mathbf{w},
	\label{io_forMP}
\end{equation} 
where $\mathbf{y}\in\mathbb{C}^{NM\times1}$ represents the received symbol vector as $\mathbf{y}(k+lN)=y[k,l]$. $\mathbf{w}\in\mathbb{C}^{NM\times1}$ denotes the additive white Gaussian noise as $\mathbf{w}(k+lN)=w[k,l]$. $\mathbf{x}\in\mathbb{C}^{N^{\prime}M^{\prime}\times1}$ stands for the sparse transmit signal as $\mathbf{x}(k+lN)=\delta[k-n_{l}]$, where $N^{\prime}=2^{N_{b}}$ denotes the available subcarrier indices. According to \eqref{io_DSK_P2P}, the sensing matrix $\mathbf{A}$ can be formulated as
\begin{equation}
	\label{OMP_A}
	\begin{aligned}
		&\mathbf{A}(k+lN,k^{\prime}+l^{\prime}N^{\prime})\\
		&=\sum_{p=1}^{P}h_{p}e^{j2\pi\frac{k_{p}l}{NM}}\delta[l-l^{\prime}-l_{p}]\delta[(k-k_{p}-k^{\prime})_{N}]
	\end{aligned}
\end{equation}
for $0\leq k\leq N-1$, $0\leq l\leq M-1$, $0\leq k^{\prime}\leq N^{\prime}-1$ and $0\leq l^{\prime}\leq M^{\prime}-1$. \par
\begin{algorithm} [t] 
	\renewcommand{\algorithmicrequire}{\textbf{Input:}}
	\renewcommand{\algorithmicensure}{\textbf{Output:}}
	\caption{MP-based detector design}
	\label{alg_MP_P2P}
	\begin{algorithmic}[1]
		\REQUIRE
		$\mathbf{y}$, $\mathbf{A}$
		\ENSURE
		$\hat{n}_{l}$ and recovered bits
		\STATE
		Initialize 
		$f_{l}=0$ for $0\leq l\leq M^{\prime}-1$ and $\mathbf{r}=\mathbf{y}$
		\FOR{$i=0:M^{\prime}-1$}
		\STATE
		$\mathbf{g}=\mathbf{A}^{H}\mathbf{r}$
		\STATE
		$\hat{q}=\arg\max_{f_{\lfloor q/N^{\prime}\rfloor}=0}|\mathbf{g}(q)|$
		\STATE
		$\mathbf{r}\leftarrow\mathbf{r}-\mathbf{a}_{\hat{q}}$
		\STATE
		$l=\lfloor\frac{\hat{q}}{N^{\prime}}\rfloor$
		\STATE
		$\hat{n}_{l}=\hat{q}-lN^{\prime}$
		\STATE
		$f_{l}=1$
		\ENDFOR
		\STATE
		Recover data bits according to $\hat{n}_{l}$
	\end{algorithmic}		
\end{algorithm}
\begin{figure*}
	\begin{equation}
		\begin{aligned}
			b_{p}[k,l]=y[(k+k_{p})_{N},l+l_{p}]-\sum_{q=1,q\ne p}^{P}h_{q}e^{j2\pi\frac{k_{q}(l+l_{p})}{NM}}\mathbb{I}_{l+l_{p}-l_{q}}\delta[(k+k_{p}-k_{q}-\hat{n}_{l+l_{p}-l_{q}})_{N}]
		\end{aligned}
		\label{SIC_P2P}
	\end{equation}
\hrulefill
\end{figure*}
\begin{figure*}
	\begin{equation}
		\begin{aligned}
			r[k,l]&=\sum_{k_{y}=0}^{N-1}z^{*}[(k_{y}-k)_{N}]y[k_{y},l]\\
			&=w_{r}[k,l]+\sum_{k_{y}=0}^{N-1}z^{*}[(k_{y}-k)_{N}]\left(\sum_{p=1}^{P}h_{p}e^{j2\pi\frac{k_{p}l}{NM}}\mathbb{I}_{l-l_{p}}x[(k_{y}-k_{p})_{N},l-l_{p}]\right)\\
			&=w_{r}[k,l]+\sum_{k_{y}=0}^{N-1}\sum_{p=1}^{P}h_{p}e^{j2\pi\frac{k_{p}l}{NM}}\mathbb{I}_{l-l_{p}}z^{*}[(k_{y}-k)_{N}]z[(k_{y}-k_{p}-n_{l-l_{p}})_{N}]\\
			&=w_{r}[k,l]+\sum_{p=1}^{P}h_{p}e^{j2\pi\frac{k_{p}l}{NM}}\mathbb{I}_{l-l_{p}}\left(\sum_{k_{y}=0}^{N-1}z^{*}[(k_{y}-k)_{N}]z[(k_{y}-k_{p}-n_{l-l_{p}})_{N}]\right)\\
			&=\sum_{p=1}^{P}h_{p}e^{j2\pi\frac{k_{p}l}{NM}}\mathbb{I}_{l-l_{p}}\delta[(k-k_{p}-n_{l-l_{p}})_{N}]+w_{r}[k,l]
		\end{aligned}
		\label{pre_sequence_DSK_derivation}
	\end{equation}
\hrulefill
\end{figure*}
Based on the formulation in \eqref{io_forMP}, the MP-based method can be utilized to execute the information recovery. $\hat{n}_{l}$ is employed to denote the estimation result of $n_{l}$ while $f_{l}\in\{0,1\}$ is utilized to mark whether $\hat{n}_{l}$ has been obtained through prior computations, i.e., $f_{l}=1$ means $\hat{n}_{l}$ has been attained. As illustrated in \textbf{Algorithm \ref{alg_MP_P2P}}, the residue vector $\mathbf{r}$ is first initialized as $\mathbf{y}$. During each iteration, the inner dot between the residue vector $\mathbf{r}$ and each column of the measured matrix $\mathbf{A}$ is computed as $\mathbf{g}$. Then the largest correlation is found from undetected multicarrier symbol indices as $\hat{q}=\arg\max_{f_{\lfloor q/N^{\prime}\rfloor}=0}|\mathbf{g}(q)|$. After that, the corresponding multicarrier symbol and subcarrier index are recovered according to $\hat{q}$. At last, the residue vector is updated by $\mathbf{r}\leftarrow\mathbf{r}-\mathbf{a}_{\hat{q}}$ considering the property of DSK modulation while $f_{l}$ is also updated. According to the formulation in \eqref{OMP_A}, for each column of $\mathbf{A}$ whose index is denoted as $k^{\prime}+l^{\prime}N^{\prime}$, there are only $P$ nonzeros lying in the $(k+lN)$-th row with $k=(k^{\prime}+k_{p})_{N}$ and $l=l^{\prime}+l_{p}$ for $p=1,2,\cdots,P$. Therefore, the complexity of each iteration can be bounded by $\mathcal{O}(M^{\prime}N^{\prime}P)$, which leads to the total complexity of less than $\mathcal{O}(M^{2}NP)$.\par 
However, according to the formulation in \eqref{OMP_A}, the nonzero elements of each column might overlap, and the corresponding coherence might be large due to the inherent sparsity of $\mathbf{A}$. For example, let we assume a simple case with $P=2$, $(l_{1},k_{1})=(0,0)$, $(l_{2},k_{2})=(1,1)$, $k_{2}=1$, $M^{\prime}=N=2$ and $M=3$. The sensing matrix $\mathbf{A}$ can be rewritten as
\begin{equation}
	\label{OMP_example}
	\mathbf{A}=\begin{pmatrix}
		h_{1}&0&0&0\\
		0&h_{1}&0&0\\
		0&h_{2}e^{j\frac{\pi}{3}}&h_{1}&0\\
		h_{2}e^{j\frac{\pi}{3}}&0&0&h_{1}\\
		0&0&0&h_{2}e^{j\frac{2\pi}{3}}\\
		0&0&h_{2}e^{j\frac{2\pi}{3}}&0
	\end{pmatrix}.
\end{equation}
The correlation coefficient between the first and fourth columns can be computed as
\begin{equation}
	\mu(\mathbf{a}_{1},\mathbf{a}_{4})=\left|\frac{\mathbf{a}_{1}^{H}\mathbf{a}_{4}}{||\mathbf{a}_{1}||_{2}||\mathbf{a}_{4}||_{2}}\right|=\frac{|h_{1}||h_{2}|}{|h_{1}|^{2}+|h_{2}|^{2}}.
\end{equation}  
It is obvious that $\mu(\mathbf{a}_{1},\mathbf{a}_{4})$ approaches $\frac{1}{2}$ when $|h_{1}|$ becomes similar to $|h_{2}|$. It would possibly cause the recovery failure of MP-based methods because of the ambiguity among columns \cite{ref_welch,ref_OMP}.
To address this issue, an iterative linear detector is proposed to enhance reliability inspired by the success of similar detectors in \cite{OTFS_MRCdetect_TVT,Lowcomdet_OTFS_ZP}. Let $c[k,l]$ and $b_{p}[k,l]$ respectively denote the detection indicator for the DSK index as $k=n_{l}$ and the corresponding indicator associated with the $p$-th path. Assuming that the coarse estimation of $n_{l}$ has been achieved from prior iterations as $\hat{n}_{l}$, $b_{p}[k,l]$ can be derived by removing the interference as \eqref{SIC_P2P} at the top of the next page. Then the indicator $c[k,l]$ can be obtained by MRC as
\begin{equation}
	\begin{aligned}
		c[k,l]=\frac{\sum_{p=1}^{P}(h_{p})^{*}e^{-j2\pi\frac{k_{p}(l+l_{p})}{NM}}b_{p}[k,l]}{\sum_{p=1}^{P}|h_{p}|^{2}}.
	\end{aligned}
\label{MRC_P2P}
\end{equation}
After that, $\hat{n}_{l}$ can be updated as
\begin{equation}
	\hat{n}_{l}=\mathop{\arg\max}_{k=0,\cdots,N^{\prime}-1}|c[k,l]|.
	\label{decision_P2P}
\end{equation}
\begin{algorithm} [t] 
	\renewcommand{\algorithmicrequire}{\textbf{Input:}}
	\renewcommand{\algorithmicensure}{\textbf{Output:}}
	\caption{Iterative SIC-MRC detector design}
	\label{alg_iter_P2P}
	\begin{algorithmic}[1]
		\REQUIRE
		$y[k,l]$, $\{h_{p},l_{p},k_{p}\}_{p=1}^{P}$
		\ENSURE
		$\hat{n}_{l}$ and recovered bits
		\STATE
		Initialize 
		$\hat{n}_{l}$
		\REPEAT
		\FOR{$l=0:M^{\prime}-1$}
		\STATE
		Compute $b_{p}[k,l]$ according to \eqref{SIC_P2P}
		\STATE
		Compute $c[k,l]$ according to \eqref{MRC_P2P}
		\STATE
		update $\hat{n}_{l}$ according to \eqref{decision_P2P}
		\ENDFOR
		\UNTIL
		stopping criteria
		\STATE
		Recover data bits according to $\hat{n}_{l}$
	\end{algorithmic}		
\end{algorithm}\par
At last, the proposed iterative SIC-MRC detector design for DSK systems is illustrated in \textbf{Algorithm \ref{alg_iter_P2P}}. After initializing\footnote{In this paper, $\hat{n}_{l}$ is initialized by executing MRC without SIC. i.e., $c[k,l]$ is first computed by utilizing \eqref{MRC_P2P} where $b_{p}[k,l]$ is substituted with $y[(k+K_{p})_{N},l+l_{p}]$. After that, \eqref{decision_P2P} is applied to initialize $\hat{n}_{l}$.} $\hat{n}_{l}$, the SIC-MRC processing is executed iteratively for $0\leq l\leq M^{\prime}-1$ to compute the indicator $c[k,l]$. Then $\hat{n}_{l}$ is updated according to \eqref{decision_P2P}. The iteration is terminated when the stopping criteria are satisfied, e.g., the maximum number of iteration times (e.g., 5) is reached and $\hat{n}_{l}$ has converged, i.e., $\hat{n}_{l}$ stays unchanged for $\forall{l}$ after a full iteration. It is worth pointing out that the iterative detector can also be applied in the flooding manner, i.e., let $\hat{n}_{l}^{(i)}$ denote the estimation results after the $i$-th iteration. In the $(i+1)$-th iteration, the SIC processing employs $\hat{n}_{l}^{(i)}$ for each $l$ rather than the updated $\hat{n}_{l}^{(i+1)}$. This enables a more efficient implementation when it comes to the computational issue. For the successive design, the computation load of each iteration can be approximately measured by $\mathcal{O}(N^{\prime}M^{\prime}P^{2})$ since the major complexity lies in the computation of $b_{p}[k,l]$. In realistic implementation, the time occupation can be further reduced if the parallel acceleration is adopted by utilizing $\hat{n}_{l}$ from prior iterations. The efficiency of the iterative SIC-MRC detector can be further improved considering that the computation of $b_{p}[k,l]$ and $c[k,l]$ can be carried out in a parallel manner for the component at each subcarrier and multicarrier symbol, where the time complexity of $\mathcal{O}(P^{2}+N^{\prime})$ is required for each iteration to update all $\hat{n}_{l}$. The performance of both parallel and successive detectors is shown in Section \ref{sec:simu}, which demonstrates the superiority beyond classic ODDM modulation systems since the parallel iteration does not lead to satisfactory decoding performance according to \cite{OTFS_MRCdetect_TVT}.\par
It is obvious that only $M^{\prime}\log{N}$ bits can be transmitted within an ODDM frame for point-to-point scenarios, which indicates a significant loss of spectral efficiency. However, by extending the DSK framework to the basis sequence-based implementation, the system throughput can increase linearly with the number of uplink UEs.\par    
\subsection{Extension to Basis Sequence-Based DSK Systems}
\label{subsec:DSK_sequence}
In this subsection, the proposed DSK framework is generalized to the basis sequence-based one, which serves as the basic of multi-user uplink transmission. Let $\{z[k]|k=0,1,\cdots,N-1\}$ represent a ZAC sequence\footnote{Considering the relation between IDFT and auto-correlation, the IDFT of $z[k]$ is a sequence with constant amplitude. It can still guarantee the low PAPR like the one-hot sequence.} with unit power, i.e., we have
\begin{equation}
	\sum_{k_{z}=0}^{N-1}z^{*}[(k_{z}-k)_{N}]z[k_{z}]=\delta[(k)_{N}].
\end{equation}
$x[k,l]$ can then be generated as 
\begin{equation}
	x[k,l]=\begin{cases}
		z[(k-n_{l})_{N}],&0\leq l\leq M^{\prime}-1\\
		0,&\text{elsewhere}
	\end{cases}.
\label{Tx_sequence_P2P}
\end{equation}
For example, the last column in Table \ref{DSK_example_table} depicts the sequence-based transmission under $N=4$.\par
\begin{figure*}
	\centering{\includegraphics[width=0.93\linewidth]{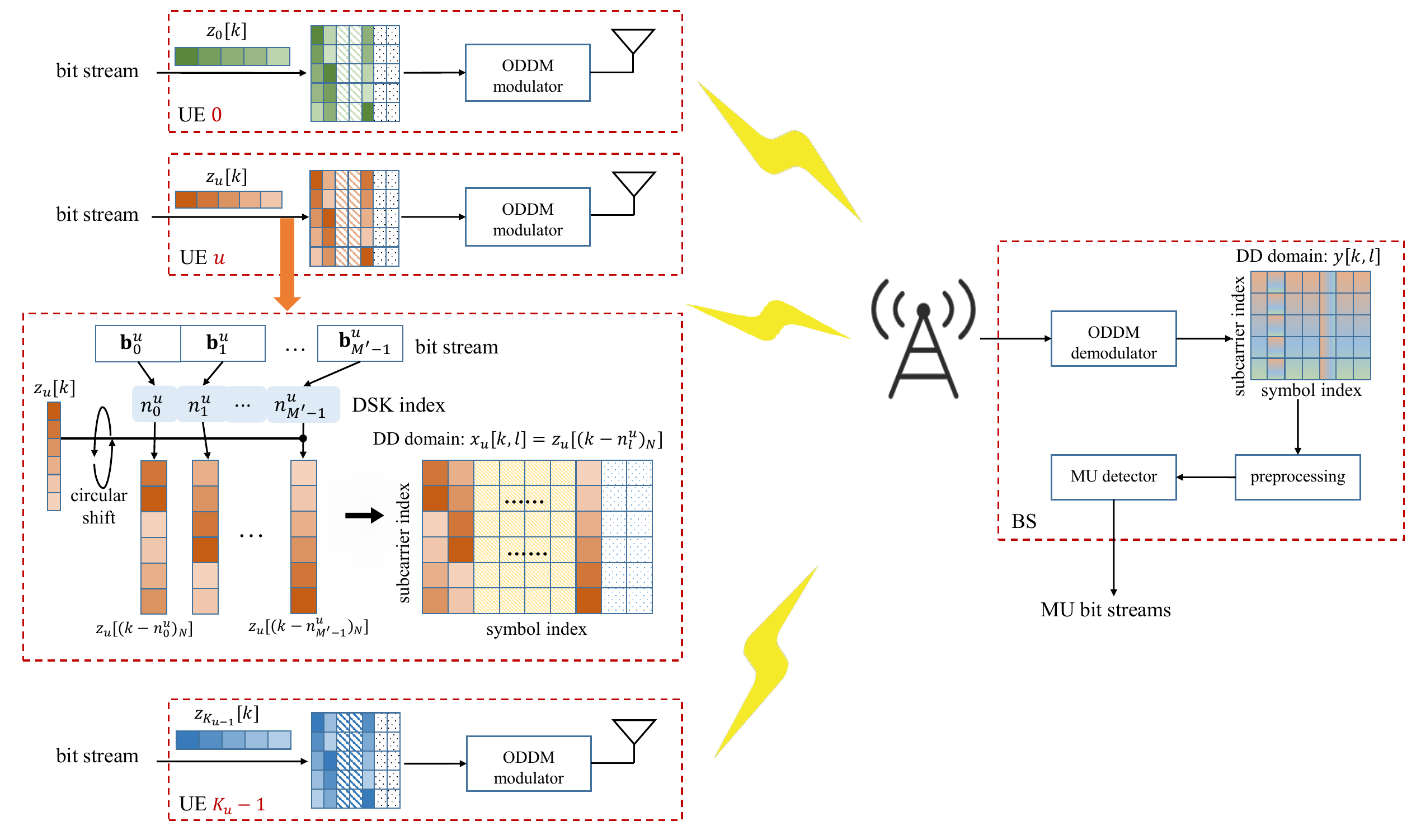}}
	\caption{Schematic of uplink DSK systems.}
	\label{Fig_system_uplink}	
\end{figure*} 
At the receiver, $y[k,l]$ is first preprocessed by employing the cross-correlation between it and the basis sequence as
\begin{equation}
	\begin{aligned}
		r[k,l]=\sum_{k_{y}=0}^{N-1}z^{*}[(k_{y}-k)_{N}]y[k_{y},l].
	\end{aligned}
\label{pre_sequence_DSK}
\end{equation}\par
Thanks to the property of $z[k]$, it can be derived as \eqref{pre_sequence_DSK_derivation} at the top of this page, where $w_{r}[k,l]\overset{\text{i.i.d}}{\sim}\mathcal{CN}(0,\sigma^{2})$ is the equivalent additive noise. It is apparent that the relation between $r[k,l]$ and $n_{l}$ stays the same as that in \eqref{io_DSK_P2P}. Therefore, both the MP-based detector in \textbf{Algorithm \ref{alg_MP_P2P}} and the iterative SIC-MRC detector in \textbf{Algorithm \ref{alg_iter_P2P}} can be utilized to recover data bits from $r[k,l]$.\par 

\section{Uplink DSK Systems with High-Mobility UEs}
\label{sec:uplink}
In this section, the discussion of point-to-point DSK systems is extended to MU uplink scenarios, where the basis sequence-based DSK implementation is adopted while each UE has a unique basis sequence as shown in Fig. \ref{Fig_system_uplink}. The iterative detector is also derived to recover data bits at BS. Finally, to enhance the reliability, the selection of basis sequences employed for DSK is discussed. The number of uplink high-mobility UEs is denoted as $K_{u}$. The impact of noise is disregarded for ease of illustration\footnote{The impact of noise is only absent for ease of the illustration since only the property of interference between the data is employed in the system design in this section. However, the noise is taken into consideration when it comes to the simulations. Future research may consider the property of noise and derive a more elaborate design.}.\par 
\subsection{Transceiver Design}
\label{subsec:MUDSK}
Let $z_{u}[k]$ denote the basis sequence for the $u$-th UE. As shown in Fig. \ref{Fig_system_uplink}, the bit stream for the $u$-th UE is first split into $M^{\prime}$ bit vectors as $\{\mathbf{b}_{l}^{u}\}_{l=0}^{M^{\prime}-1}$. $\mathbf{b}_{l}^{u}$ is then transferred into the DSK index as $n_{l}^{u}$. After that, data components for the $u$-th UE can be generated like that in \eqref{Tx_sequence_P2P} as
\begin{equation}
	x_{u}[k,l]=\begin{cases}
		z_{u}[(k-n_{l}^{u})_{N}],&0\leq l\leq M^{\prime}-1\\
		0,&\text{elsewhere}
	\end{cases},
\end{equation}
which leads to the baseband waveform $s_{u}(t)$ by utilizing the ODDM modulator.\par 
At the BS side, the ODDM demodulator is first employed to recover demodulated components as $y[k,l]$. Let $h_{p}^{u}$, $l_{p}^{u}$, $k_{p}^{u}$ denote the channel parameters associated with the $p$-th path for the $u$-th UE, respectively. According to \eqref{DDIO_final}, it can be derived as
\begin{equation}
	\begin{aligned}
		y[k,l]=\sum_{u=0}^{K_{u}-1}\left(\sum_{p=1}^{P_{u}}h_{p}^{u}e^{j2\pi\frac{k_{p}^{u}l}{NM}}\mathbb{I}_{l-l_{p}^{u}}z_{u}[(k-k_{p}^{u}-n_{l-l_{p}^{u}}^{u})_{N}]\right)
	\end{aligned},
\end{equation}
where $P_{u}$ represents the number of paths of the $u$-th UE. Let $\psi_{u,v}(k_{u},k_{v})$ denote the cross-correlation\footnote{Thanks to the property of circular shift, $\psi_{u,v}(k_{u},k_{v})$ is only related to $u$, $v$ and $(k_{u}-k_{v})_{N}$, which benefits the practical storage complexity. In this paper, $\psi_{u,v}(k_{u},k_{v})$ is adopted only for ease of illustration.} of basis sequences between the $u$-th and $v$-th UE, which is defined as
\begin{equation}
	\psi_{u,v}(k_{u},k_{v})=\sum_{k=0}^{N-1}z_{u}^{*}[(k-k_{u})_{N}]z_{v}[(k-k_{v})_{N}],
\end{equation}
where $\psi_{u,u}(0,0)=1$ is required to normalize the power. After similar preprocessing like \eqref{pre_sequence_DSK}, $r_{u}[k,l]$ can be obtained as
\begin{equation}
	\begin{aligned}
		&r_{u}[k,l]=\sum_{k_{y}=0}^{N-1}z_{u}^{*}[(k_{y}-k)_{N}]y[k_{y},l]\\
		&=\sum_{k_{y}=0}^{N-1}z_{u}^{*}[(k_{y}-k)_{N}]\\
		&\times\sum_{v=0}^{K_{u}-1}\sum_{p=1}^{P_{v}}h_{p}^{v}e^{j2\pi\frac{k_{p}^{v}l}{NM}}\mathbb{I}_{l-l_{p}^{v}}z_{v}[(k_{y}-k_{p}^{v}-n_{l-l_{p}^{v}}^{v})_{N}]\\
		&=\sum_{v=0}^{K_{u}-1}\sum_{p=1}^{P_{v}}h_{p}^{v}e^{j2\pi\frac{k_{p}^{v}l}{NM}}\mathbb{I}_{l-l_{p}^{v}}\\
		&\times\sum_{k_{y}=0}^{N-1}z_{u}^{*}[(k_{y}-k)_{N}]z_{v}[(k_{y}-k_{p}^{v}-n_{l-l_{p}^{v}}^{v})_{N}]\\
		&=\sum_{v=0}^{K_{u}-1}\sum_{p=1}^{P_{v}}h_{p}^{v}e^{j2\pi\frac{k_{p}^{v}l}{NM}}\mathbb{I}_{l-l_{p}^{v}}\psi_{u,v}(k,(k_{p}^{v}+n_{l-l_{p}^{v}}^{v})_{N}).
	\end{aligned}
	\label{pre_DSK_uplink}
\end{equation}
Similar to the analysis in Section \ref{sec:P2P} around \eqref{OMP_example}, the correlation among columns of the sensing matrix might be strong, which harms the performance of MP-based methods. Therefore, we consider employing the iterative detector instead, which is illustrated in \textbf{Algorithm \ref{alg_iter_uplink}}. Assuming that $\psi_{u,u}(k_{1},k_{2})=\delta[(k_{1}-k_{2})_{N}]$ holds for all UEs.  At first, the estimation of $n_{l}^{u}$ is initialized\footnote{Considering the computation load of the initialization, it is treated as one iteration during the following contents including the simulations.} as
\begin{equation}
	\begin{aligned}
		(\hat{n}_{l}^{u})^{(1)}&=\mathop{\arg\max}_{k=0,1,\cdots,N^{\prime}-1}|c_{u}[k,l]|,\\
		c_{u}[k,l]&=\frac{\sum_{p=1}^{P_{u}}(h_{p}^{u})^{*}e^{-j2\pi\frac{k_{p}^{u}(l+l_{p}^{u})}{NM} }r_{u}[(k+k_{p}^{u})_{N},l+l_{p}^{u}]}{\sum_{p=1}^{P_{u}}|h_{p}^{u}|^{2}},
	\end{aligned}
\label{initial_uplink}
\end{equation}
where $c_{u}[k,l]$ serves as the MRC indicator for $n_{l}^{u}=k$ like the point-to-point scenarios. Let $(\hat{n}_{l}^{u})^{(i)}$ represent the detection results from the $i$-th iteration. During the $(i+1)$-th iteration, since the indicator for $n_{l}^{u}=k$ is only related to $P_{u}$ variables as $\{r_{u}[(k+k_{p}^{u})_{N},l+l_{p}^{u}]\}_{p=1}^{P_{u}}$, the corresponding $P_{u}$ elements $\{b_{p}^{u}[k,l]\}_{p=1}^{P_{u}}$ for $n_{l}^{u}=k$ can be derived by removing the interference as \eqref{SIC_uplink} at the top of the next page, which is deduced according to the characterization in \eqref{pre_DSK_uplink}.
\begin{figure*}
\begin{equation}
	\begin{aligned}
		b_{p}^{u}[k,l]&=r_{u}[(k+k_{p}^{u})_{N},l+l_{p}^{u}]-\sum_{q=1,q\ne p}^{P_{u}}\mathbb{I}_{l+l_{p}^{u}-l_{q}^{u}}h_{q}^{u}e^{j2\pi\frac{k_{q}^{u}(l+l_{p}^{u})}{NM}}\psi_{u,u}\left((k+k_{p}^{u})_{N},\left(k_{q}^{u}+(\hat{n}^{u}_{l+l_{p}^{u}-l_{q}^{u}})^{(i)}\right)_{N}\right)\\
		&-\sum_{v=0,v\ne u}^{K_{u}-1}\sum_{q=1}^{P_{v}}\mathbb{I}_{l+l_{p}^{u}-l_{q}^{v}}h_{q}^{v}e^{j2\pi\frac{k_{q}^{v}(l+l_{p}^{u})}{NM}}\psi_{u,v}\left((k+k_{p}^{u})_{N},\left(k_{q}^{v}+(\hat{n}^{v}_{l+l_{p}^{u}-l_{q}^{v}})^{(i)}\right)_{N}\right)
	\end{aligned}
\label{SIC_uplink}
\end{equation}
\hrulefill
\end{figure*}
After that, these $P_{u}$ variables are combined under the MRC rule as
\begin{equation}
	c_{u}[k,l]=\frac{\sum_{p=1}^{P_{u}}(h_{p}^{u})^{*}e^{-j2\pi\frac{k_{p}^{u}(l+l_{p}^{u})}{NM} }b_{p}^{u}[k,l]}{\sum_{p=1}^{P_{u}}|h_{p}^{u}|^{2}},
	\label{MRC_uplink}
\end{equation}
which serves as the final indicator for whether $k=n_{l}^{u}$ holds. In another word, the estimation of $n_{l}^{u}$ is updated as
\begin{equation}
	(\hat{n}_{l}^{u})^{(i+1)}=\mathop{\arg\max}_{k=0,1,\cdots,N^{\prime}-1}|c_{u}[k,l]|.
	\label{decision_uplink}
\end{equation}
\begin{algorithm} [t] 
	\renewcommand{\algorithmicrequire}{\textbf{Input:}}
	\renewcommand{\algorithmicensure}{\textbf{Output:}}
	\caption{Iterative SIC-MRC detector with parallel manner for uplink DSK systems}
	\label{alg_iter_uplink}
	\begin{algorithmic}[1]
		\REQUIRE
		$r_{u}[k,l]$, relevant channel parameters
		\ENSURE
		$\hat{n}_{l}^{u}$ and recovered bits
		\STATE
		Initialize 
		$(\hat{n}_{l}^{u})^{(1)}$ according to \eqref{initial_uplink}
		\STATE
		Initialize iteration number as $i=1$
		\REPEAT
		\FOR{$l=0:M^{\prime}-1$}
		\FOR{$u=0:K_{u}-1$}
		\STATE
		Compute $b_{p}^{u}[k,l]$ according to \eqref{SIC_uplink}
		\STATE
		Compute $c_{u}[k,l]$ according to \eqref{MRC_uplink}
		\STATE
		update $(\hat{n}_{l}^{u})^{(i+1)}$ according to \eqref{decision_uplink}
		\ENDFOR
		\ENDFOR
		\STATE
		$i\leftarrow i+1$
		\UNTIL
		stopping criteria
		\STATE
		Let $\hat{n}_{l}^{u}=(\hat{n}_{l}^{u})^{(i)}$ and recover data bits according to $\hat{n}_{l}^{u}$
	\end{algorithmic}		
\end{algorithm}\par 
The iteration is terminated when the stopping criteria are satisfied, e.g., the maximum number of iterations (e.g., 10) has been reached or the results of data recovery have converged, i.e., $(\hat{n}_{l}^{u})^{(i)}=(\hat{n}_{l}^{u})^{(i-1)}$ for $\forall l,u$. It is worth pointing out that the proposed scheme in \textbf{Algorithm \ref{alg_iter_uplink}} reveals the parallel manner, i.e., the update of $(\hat{n}_{l}^{u})^{(i+1)}$ is uncorrelated with each other for $l$ and $u$, which can reduce the complexity significantly. Considering the parallel execution, the total complexity\footnote{This is deduced because the major complexity lies in \eqref{SIC_uplink} and \eqref{decision_uplink}.} can be bounded as $\mathcal{O}\left(N_{\text{iter}}(P_{\text{s}}+N^{\prime})\right)$, where $P_{\text{s}}=\sum_{u=0}^{K_{u}-1}P_{u}$ denotes the total number of MU channel paths while $N_{\text{iter}}$ stands for the maximum iteration times. For conventional ODDM systems with the same spectral efficiency, the typical LMMSE and MPA-based detection schemes require the complexity order of $\mathcal{O}\left((K_{u}N_{b}M^{\prime})^{3}\right)$ and $\mathcal{O}\left(N_{\text{iter}}K_{u}N_{b}P_{s}M^{\prime}\right)$, respectively. Therefore, the DSK framework with the proposed parallel SIC-MRC detection reveals much lower complexity overhead since $N^{\prime}<M^{\prime}$ is satisfied in most of the ODDM systems \cite{ref_ODDM,ODDM_tutorial,ODDM_precoding}. Future research might consider the combination of the DSK modulation and conventional detection designs in ODDM systems to realize a better complexity-performance trade-off.\par 
\subsection{Selection of Basis Sequences}
\label{subsec:ZC}
In this subsection, the selection of basis sequences for MU uplink DSK systems is discussed. Considering the characterization in \eqref{pre_DSK_uplink}, sequences satisfying $\psi_{u,v}(k_{u},k_{v})=\delta[u-v]\delta[(k_{u}-k_{v})_{N}]$ are ideal for the DSK implementation. However, this cannot be achieved since the maximum number of standard orthogonal vectors in $\mathbb{C}^{N}$ is $N$, which is far less than $K_{u}N^{\prime}$ as long as $K_{u}\geq2$. As a result, we can only minimize values of $|\psi_{u,v}(k_{u},k_{v})|$ for $u\ne v$ and $k_{u}\ne k_{v}$ rather than force them to zeros. On the other hand, if $\psi_{u,u}(k_{1},k_{2})$ has nonzero for $(k_{1}-k_{2})_{N}\ne0$, $c_{u}[k,l]$ cannot serve as the qualified indicator for $k=n_{l}^{u}$. As a result, the basis sequences are required to satisfy $\psi_{u,u}(k_{1},k_{2})=\delta[(k_{1}-k_{2})_{N}]$ at first, i.e., $z_{u}[k]$ must be a ZAC sequence for $\forall~u$. Then basis sequences are selected to minimize the maximum cross-correlation as $\psi_{\text{max}}=\max_{u\ne v,0\leq k_{u},k_{v}\leq N^{\prime}-1}|\psi_{u,v}(k_{u},k_{v})|$. Let $\psi_{\text{max}}^{(u,v)}=\max_{0\leq k_{u},k_{v}\leq N^{\prime}-1}|\psi_{u,v}(k_{u},k_{v})|$ denote the maximum value of cross-correlation between the $u$-th UE and $v$-th UE for $u\ne v$. If two choices share the same value of $\psi_{\text{max}}$, the number of UE pairs $(u,v)$ with $\psi_{\text{max}}^{(u,v)}=\psi_{\text{max}}$ should be minimized.\par 
Considering the aforementioned analysis, ZC sequences \cite{ref_ZC_chu} with different roots are appropriate to serve as the basis sequences, which can be derived as
\begin{equation}
	z_{u}[k]=\frac{1}{\sqrt{N}}\times\begin{cases}
		e^{j\pi m_{u}\frac{k^{2}}{N}},&N \text{ is even}\\
		e^{j\pi m_{u}\frac{k(k+1)}{N}},&N \text{ is odd}
	\end{cases},
\label{ZC}
\end{equation}
where $0<m_{u}<N$ is the root allocated to the $u$-th UE. To satisfy the ZAC property, $m_{u}$ is required to be relatively prime to $N$ according to \cite{ref_ZC_chu}. Then the selection of sequences is translated into the optimization of $m_{u}$ for $u=0,\cdots,K_{u}-1$. Since the number of data bits associated with per multicarrier symbol index is $N_{b}=\lfloor\log_{2}N\rfloor$, we focus\footnote{If $N$ is a prime number, MU roots can be selected randomly. The investigation of other values of $N$ may further enhance the practicality, which will be embodied in our future researches.} on the analysis of $N=2^{n}$ for $n\in\mathbb{N}^{+}$ to maximize the utilization efficiency of DD domain subcarriers. Different from the analysis in \cite{ref_ZC_cross}, a more simplified derivation is provided in the following lemma about $\psi_{u,v}(k_{u},k_{v})$ to serve as the basis of the discussion.
\begin{lemma}
	\rm
	\label{lemma1_cross_correlation}
	$|\psi_{u,v}(k_{u},k_{v})|^{2}$ can be derived as
	\begin{equation}
		\begin{aligned}
			&|\psi_{u,v}(k_{u},k_{v})|^{2}\\
			&=\begin{cases}
				\frac{c}{N},&\left((k_{u}m_{u}-k_{v}m_{v})_{N}\right)_{c}=0\\
				0,&\text{elsewhere}	
			\end{cases},
		\end{aligned}
	\end{equation}
where we have $c=\text{gcd}(m_{v}-m_{u},N)$.
	\begin{IEEEproof}
		The proof is provided in Appendix \ref{lemma1_proof}.
	\end{IEEEproof}
\end{lemma}\par
Based on the conclusions of \textbf{Lemma \ref{lemma1_cross_correlation}}, $\psi_{\text{max}}$ is determined by the maximum value of $\text{gcd}(m_{v}-m_{u},N)$. Therefore, the proposed optimization criteria can be transferred into minimizing the maximum value of $\text{gcd}(m_{v}-m_{u},N)$ and the number of accessible pairs. Without loss of generality, we assume that $0<m_{0}<m_{1}<\cdots<m_{K_{u}-1}$ holds. Then the following theorem can be derived.
\begin{theorem}
	\rm
	\label{theorem_MUsequences}
	The optimal selection of $m_{u}$ under the proposed criteria is $m_{u}=2u+m_{0}$, i.e., the sequence $\{m_{u}|u=0,1,\cdots,K_{u}-1\}$ forms an arithmetic sequence with a common difference of $2$. $m_{0}$ must be an odd number to satisfy $\text{gcd}(m_{u},N)=1$ for $\forall~u$.
	\begin{IEEEproof}
		The proof is provided in Appendix \ref{theorem_MUsequences_proof}.
	\end{IEEEproof}
\end{theorem}\par
\textbf{Theorem \ref{theorem_MUsequences}} offers the design of MU basis sequences by providing the optimal roots allocated to each UE, which completes the framework of uplink DSK systems. Nevertheless, the proposed allocation might lead to an imbalance among UEs since different pairs of UEs might bear different levels of interference denoted as $\frac{\text{gcd}(m_{v}-m_{u},N)}{N}=\frac{\text{gcd}(2(v-u),N)}{N}$ according to the analysis in \textbf{Theorem \ref{theorem_MUsequences}} and \textbf{Lemma \ref{lemma1_cross_correlation}}. For example, let we consider a simple case with $3$ UEs and $m_{0}=1$, the peak interference power between the first and third UEs can be computed as $\frac{4}{N}$ while that between the first and second UEs is $\frac{2}{N}$. It can then be easily deduced that the second UE bears a lower interference level. This imbalance might possibly lead to different performance among UEs. Further system optimization and evaluation considering the imbalance between UEs requires a more elaborate system design, which falls outside of the scope of this paper and will be considered in our future work.\par 
Thanks to the MUI mitigation enabled by the optimization of MU basis sequences in \textbf{Theorem \ref{theorem_MUsequences}}, the overall throughput for uplink DSK systems increases linearly\footnote{Further enhancement of the spectral efficiency can be expected by involving the techniques of index modulation with practical alphabets, i.e., let $x_{u}[k,l]=\alpha_{l}^{u}z_{u}[(k-n_{l}^{u})_{N}]$ denote the transmission frame in the DD domain with $\alpha_{l}^{u}$ employed to load data bits, which will certainly be embodied in our future research.} with the number of UEs while the BER performance remains almost the same as that in point-to-point transmission as illustrated in Section \ref{sec:simu}. The number of data bits transmitted within an ODDM frame can be derived by $K_{u}M^{\prime}\log_{2}{N}$, while that of the binary phase shift keying (BPSK)-enabled ODDM system is less than $M^{\prime}N$ considering the guard interval reserved for mitigating MUI \cite{MA_OTFS_simu_WCL}. Therefore, the proposed DSK framework maintains similar transmission efficiency to conventional ODDM systems when $K_{u}$ approaches $\frac{N}{\log_{2}{N}}$, e.g., 10 for $N=64$. On the other hand, the PAPR of the DSK signal is far lower than BPSK systems thanks to the constant modulus of the IFFT of the ZAC sequence, while the time complexity of the detection is also lower than conventional ODDM systems considering the parallel iterative SIC-MRC scheme in \textbf{Algorithm \ref{alg_iter_uplink}}. Therefore, it is obvious that the proposed DSK framework can serve as a qualified candidate for uplink transmission over doubly dispersive channels.
\par 
\section{Performance Evaluation}
\label{sec:simu}
\begin{table}
	\caption{Simulation Parameters}
	\centering
	\label{simulation_para_table}
	\renewcommand\arraystretch{1.3}
	\begin{tabular}{p{16em}|p{10em}}
		\hline
		Parameter &
		Typical value\\
		\hline
		Carrier frequency ($f_{c}$)& $5$ GHz\\
		Number of subcarriers ($N$)& $64$\\
		Number of multicarrier symbols ($M$)& $256$\\
		Sampling frequency ($f_{s}=1/\tau_{r}$)& $3840=256\times15$ kHz\\
		UE speed (km/h)& $360$,~$500$\\
		Maximum delay index of channel ($l_{\text{max}}$)& $10$\\
		Number of channel taps & $4,~5$\\
		\hline
	\end{tabular}
\end{table}
In this section, the performance of the proposed DSK modulation is evaluated by simulation results. Typical values of relevant parameters are provided in Table \ref{simulation_para_table} similar to those in \cite{ref_ODDM,ODDM_cheng_simu_tcom}. Following the settings in \cite{ODDM_cheng_simu_tcom}, the Doppler and delay index of each path are randomly selected from $[-k_{\text{max}},k_{\text{max}}]$ and $[0,l_{\text{max}}]$ for each channel realization, where $k_{\text{max}}$ is determined by the maximum mobility of UEs. For example, $k_{\text{max}}=8$ can be derived with the maximum mobility speed as $v_{\text{max}}=360$ km/h. The complex gain of each path is randomly generated as $h_{p}^{u}\overset{\text{i.i.d.}}{\sim}\mathcal{CN}(0,\frac{1}{P_{u}})$. The signal-to-noise ratio (SNR) is defined as $\rho=\frac{1}{N_{b}\sigma^{2}}$ since basis sequences have been normalized as in \eqref{ZC}. At last, simulations of DSK are based on the ZC sequences-based implementation.\par
\subsection{Point-to-Point Scenarios}
\label{subsec:simu_P2P}
\begin{figure}
	\centering{\includegraphics[width=0.83\linewidth]{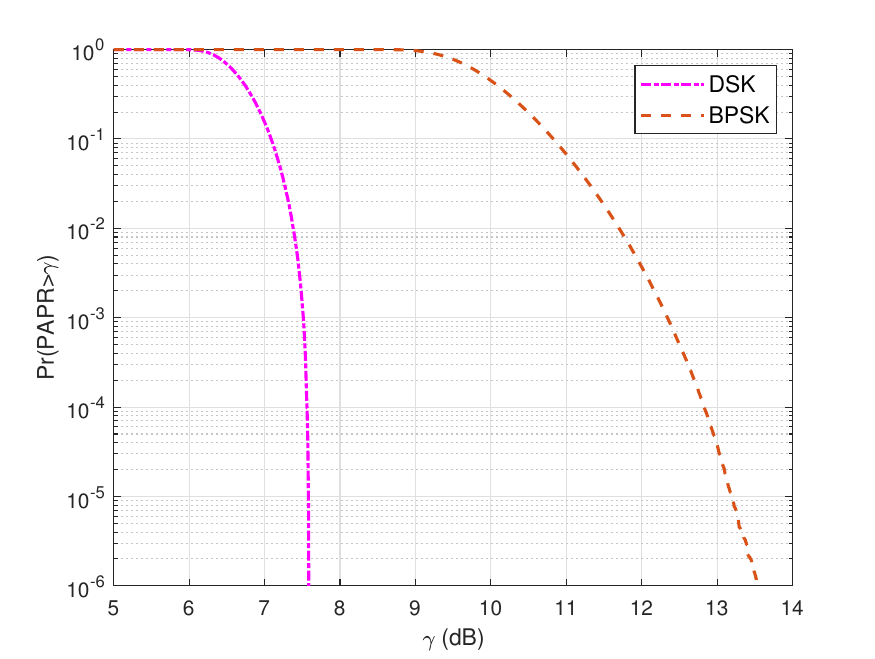}}
	\caption{CCDF comparison for DSK and BPSK-enabled DD domain modulation systems.}
	\label{SimuFig_PAPR}	
\end{figure}
The performance is first evaluated for point-to-point scenarios, where the traditional DD domain framework with BPSK alphabets serves as the comparison baseline. For BPSK-based ODDM schemes, the interleaved pattern is adopted to reduce the interference and keep consistent with the spectral efficiency in DSK systems, where $N_{d}$ denotes the number of null subcarriers between 2 adjacent BPSK symbols within each ODDM multicarrier symbol. In another word, BPSK symbols are inserted at $\{x[k,l]|l=0,1,\cdots, M^{\prime}-1,k=i(N_{d}+1),i=0,1,\cdots, N_{b}-1\}$, where we have $N_{b}=\lfloor\log_{2}N\rfloor=6$ and $N_{d}=\lfloor\frac{N}{N_{b}}\rfloor-1=9$ considering the parameter values in Table \ref{simulation_para_table}. The other null subcarriers are treated as the guard interval to reduce the interference between data components, which can also promote the BER of BPSK-enabled ODDM systems. The LMMSE detector in \cite{OTFS_window_LMMSE} is performed at the receiver for BPSK-enabled ODDM systems. \par 
The complementary cumulative distribution function (CCDF) comparison is first presented in Fig. \ref{SimuFig_PAPR} to show the superiority of our proposed DSK systems. The PAPR $\gamma$ is approximated by oversampling $s(t)$ with $8f_{s}$, where the raised cosine filter is adopted as the pulse-shaping with roll-off factor $\alpha=0.1$. PAPR is not $0$ dB even though the ZAC sequence is adopted due to ZP and pulse-shaping, however, the proposed DSK modulation still outperforms the BPSK scheme significantly. About $5$ dB reduction can be observed for the CCDF threshold as $10^{-3}$. The CCDF value of $\gamma=7.6$ dB is about $10^{-5}$ for DSK-ODDM systems, however, the PAPR of BPSK-ODDM schemes is almost always higher than $8$ dB. As a result, we conclude that DSK is a better implementation technique of DD domain transmission when it comes to PAPR issues.\par 
\begin{figure}
	\centering{\includegraphics[width=0.83\linewidth]{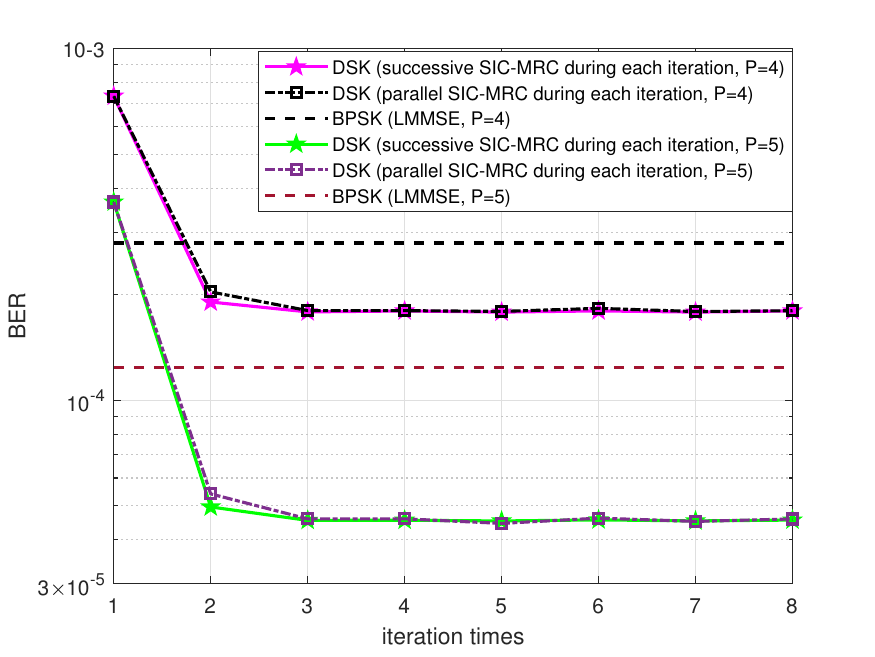}}
	\caption{BER against iteration times under $\rho=12$ dB and $v_{\text{max}}=360$ km/h.}
	\label{SimuFig_SU_iter}	
\end{figure}
The convergence of the proposed iterative SIC-MRC detector in \textbf{Algorithm \ref{alg_iter_P2P}} is then verified in Fig. \ref{SimuFig_SU_iter} by plotting BER performance against iterations under $\rho=12$ dB and $v_{\text{max}}=360$ km/h. The performance is promoted with the number of paths $P$ increasing, which is consistent with results in \cite{OTFS_performance_P_TWC,OTFS_iterdetect_P_TWC}. This is because the potential multipath diversity increases with the number of paths, which helps enhance the reliability. After two iteration times, the BER of our proposed detector can outperform that of BPSK-enabled ODDM modulation systems with the LMMSE\footnote{We consider the LMMSE detector since it enjoys no error propagation \cite{ODDM_precoding,Lowcomdet_ODDM_ZP}. However, the proposed DSK framework does not have LMMSE detection due to the non-linear shift keying operation.} detector. BER of our proposed scheme converges to less than $2\times10^{-4}$ under $P=4$ after about $5$ iterations, which is reduced to less than $5\times10^{-5}$ with $P=5$. On the other hand, the parallel detector almost approaches the performance of the successive design, which indicates the significant computational superiority of our proposed scheme. \par 
\begin{figure}
	\centering{\includegraphics[width=0.83\linewidth]{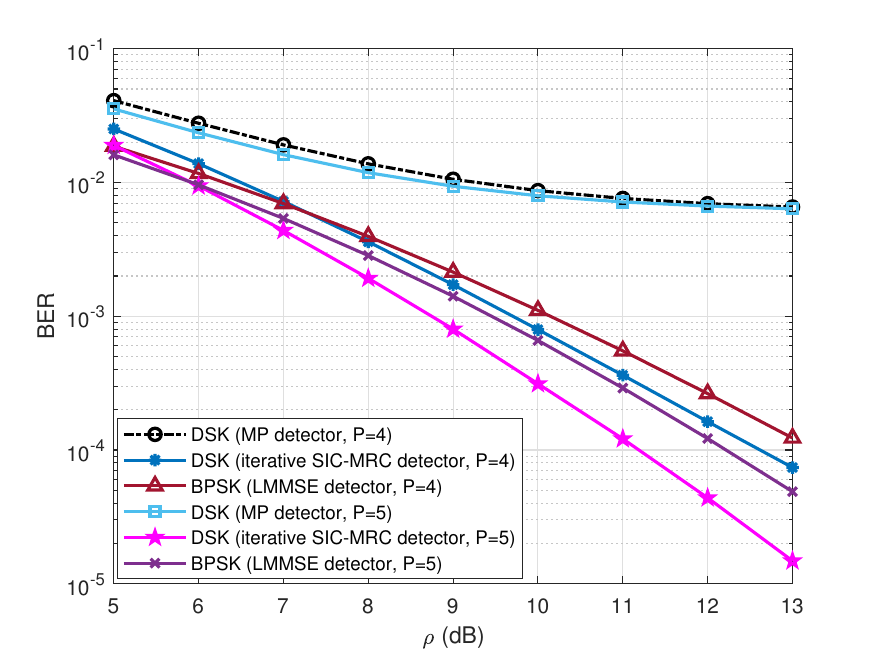}}
	\caption{BER against $\rho$ under $v_{\text{max}}=360$ km/h.}
	\label{SimuFig_SU_SNR}	
\end{figure}
Fig. \ref{SimuFig_SU_SNR} depicts the BER curve against SNR under $v_{\text{max}}=360$ km/h, with a maximum of $5$ iterations for the iterative SIC-MRC detector. When $\rho=11$ dB, BER of DSK with the iterative detector approaches $1\times10^{-4}$ with $P=5$. However, the BER of the BPSK system is about $3\times10^{-4}$, which verifies the superiority of the proposed DSK framework. Under $4$ paths, DSK outperforms BPSK by more than $0.2$ dB when BER is about $10^{-3}$, which is amplified to more than $1$ dB if $5$ paths are involved. Finally, the BER of the MP detector is the highest regardless of the SNR value, which caters to the analysis in Section \ref{sec:P2P}. On the other hand, DSK with the iterative detector provides the lowest complexity and BER among the simulated schemes, which demonstrates the value of our work again.\par      
\subsection{Uplink Scenarios}
\label{subsec:simu_uplink}
In this subsection, the performance is evaluated under uplink scenarios with high-mobility UEs. For the comparison baseline, distinct UEs are allocated with non-overlapping ODDM subcarriers for UEs like \cite{MA_OTFS_simu_WCL,MA_OTFS_simu_Tcom}, while the subcarriers for each UE form a subband. It can be treated as a frequency domain multiple access (FDMA) strategy with respect to the Doppler resolution. To be more specific, let $N_{s}=\lfloor\frac{N}{K_{u}}\rfloor$ denote the number of subcarriers allocated to each UE. BPSK symbols of the $u$-th UE are inserted at $k=uN_{s},uN_{s}+1,\cdots,uN_{s}+N_{b}-1$ for $0\leq l\leq M^{\prime}-1$ with $N_{s}\geq N_{b}$. The other null subcarriers serve as the guard interval to mitigate the MU interference (MUI) between distinct subcarriers. Therefore, $K_{u}N_{b}M^{\prime}$ bits are loaded within each ODDM frame for both DSK and BPSK schemes throughout the simulations. At the BS side, the LMMSE detector is deployed similarly to \cite{MA_OTFS_simu_Tcom,OTFS_MA_Tcom_LMMSE_ZP} for uplink BPSK-ODDM systems. All the simulation results in this subsection are averaged over more than $10000$ ODDM frames. \par 
\begin{figure}
	\centering{\includegraphics[width=0.83\linewidth]{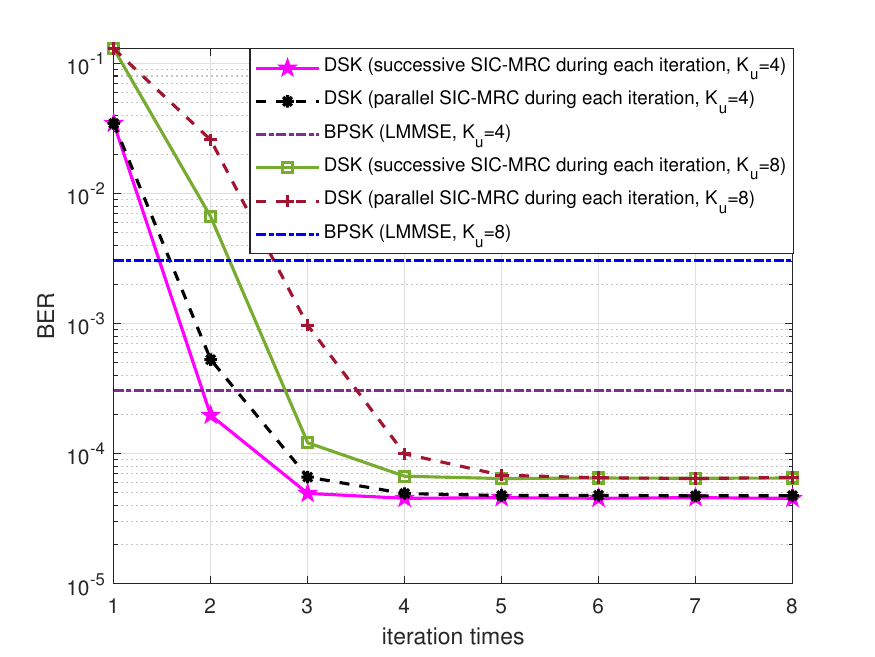}}
	\caption{BER against iteration times under $\rho=12$ dB, $P_{u}=5$ and $v_{\text{max}}=360$ km/h.}
	\label{SimuFig_MU_iter}	
\end{figure}
At first, the convergence test is executed in Fig. \ref{SimuFig_MU_iter} under $\rho=12$ dB, $P_{u}=5$ and $v_{\text{max}}=360$ km/h to verify the reliability of the proposed iterative SIC-MRC detector in \textbf{Algorithm \ref{alg_iter_uplink}}. Compared with point-to-point scenarios, the convergence rate is slower due to the existence of MUI. However, an excellent BER performance can be attained for both parallel and successive iterative detectors after about $5$ iterations, where BER is less than $7\times10^{-5}$ under $K_{u}=8$. However, the BER of the BPSK-enabled DD domain uplink scheme is more than $2\times10^{-3}$ even though the optimal LMMSE detector can be deployed with $K_{u}=8$, which demonstrates the superiority of the proposed DSK framework. Meanwhile, the BER increases significantly for BPSK-based schemes, where the BER of $K_{u}=8$ is ten times as much as that of $K_{u}=4$. Thanks to the iterative SIC-MRC detector and the optimization of MU basis sequences, our proposed DSK framework reveals better robustness with $K_{u}$ increasing, where the BER increment is only about $2\times10^{-5}$ if $K_{u}$ increases from $4$ to $8$. Finally, since an excellent BER performance can be achieved after $5$ iterations for any scenario, the maximum number of iterative times is set as $5$ in the following simulations.\par  
\begin{figure}
	\centering{\includegraphics[width=0.83\linewidth]{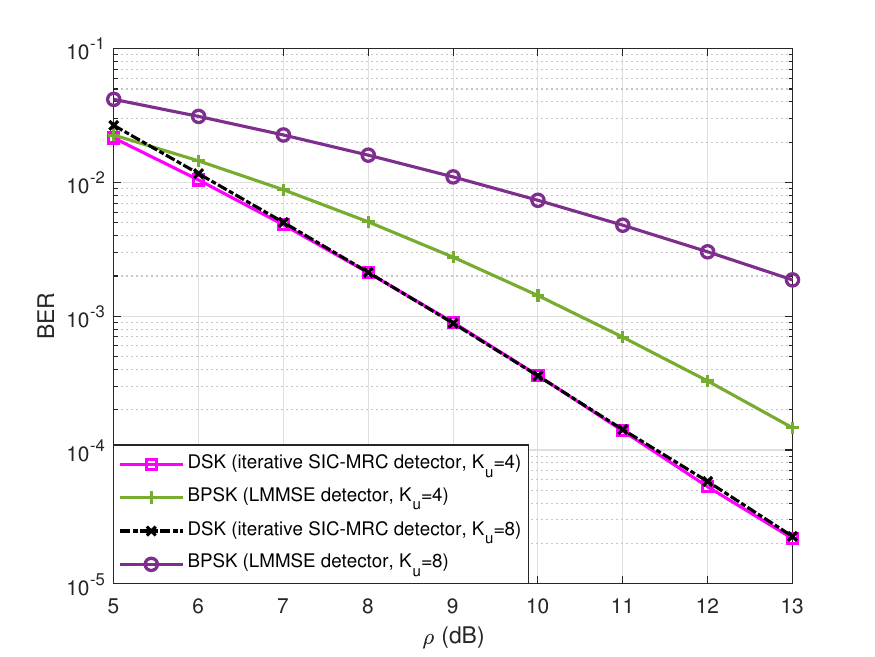}}
	\caption{BER against $\rho$ under $P_{u}=5$ and $v_{\text{max}}=360$ km/h.}
	\label{SimuFig_MU_SNR}	
\end{figure}
Fig. \ref{SimuFig_MU_SNR} illustrates the BER performance against SNR, where we set $P_{u}=5$ and $v_{\text{max}}=360$ km/h. It is apparent that the proposed DSK framework demonstrates better robustness with $K_{u}$ increasing. When $\rho=11$ dB, BER of the proposed DSK scheme approaches $10^{-4}$ under $8$ uplink UEs, which is less than $3\%$ of that in BPSK-enabled DD domain systems. When BER meets $10^{-3}$, DSK outperforms BPSK by about $1.5$ dB with $4$ uplink UEs, which is amplified to more than $4$ dB if $K_{u}=8$. Therefore, DSK can guarantee a much more reliable performance than BPSK-based DD domain systems, especially for MU uplink scenarios, which is essential in realistic high-mobility applications.\par 
\begin{figure}
	\centering{\includegraphics[width=0.83\linewidth]{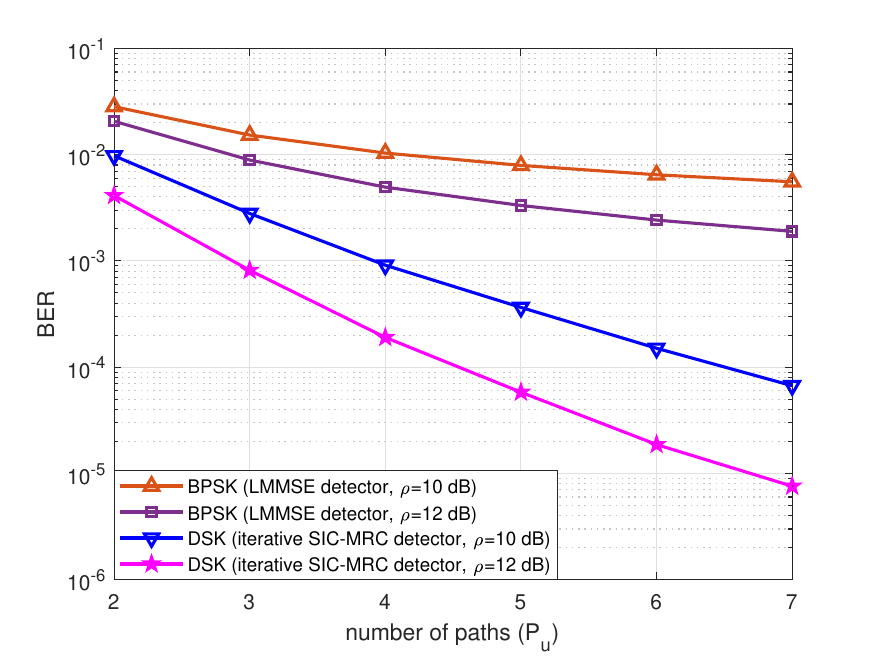}}
	\caption{BER against $P_{u}$ under $K_{u}=8$.}
	\label{SimuFig_MU_Pu}	
\end{figure}
The performance precedence is then verified by plotting BER against the number of paths $P_{u}$ in Fig. \ref{SimuFig_MU_Pu}, where we set $K_{u}=8$ and $v_{\text{max}}=500$ km/h. BER decreases with $P_{u}$ increasing thanks to the diversity gain of multipath channels, where the performance advantage of the DSK scheme is also amplified. When there are $6$ incident paths for each UE, the BER of DSK approaches $10^{-4}$ with $\rho=10$ dB, which is less than $2.3\%$ of that in BPSK systems with the same conditions. BER of the proposed DSK framework is less than $10^{-5}$ when $\rho=12$ dB and $P_{u}=7$. Nevertheless, the BER of BPSK-based ODDM modulation systems is still more than $10^{-3}$ under these conditions, which demonstrates the value of the proposed DSK modulation again.\par  
\begin{figure}
	\centering{\includegraphics[width=0.83\linewidth]{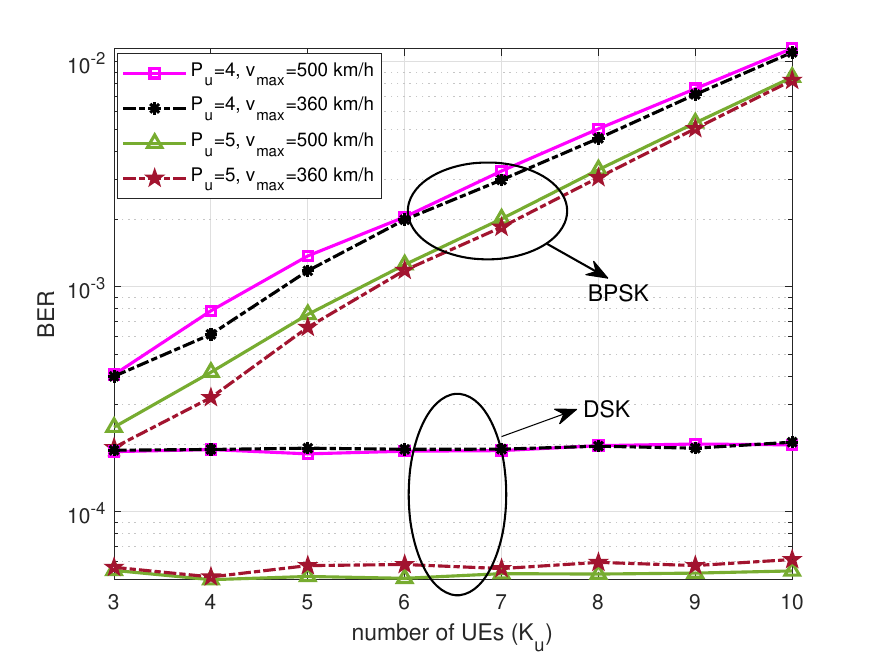}}
	\caption{BER against $K_{u}$ under $\rho=12$ dB.}
	\label{SimuFig_MU_Ku}	
\end{figure}
The BER performance against $K_{u}$ is depicted in Fig. \ref{SimuFig_MU_Ku} under $\rho=12$ dB, where the influences of $P_{u}$ and $v_{\text{max}}$ are also embodied. It is obvious that there is only a very slight performance difference as the maximum velocity increases. It demonstrated that the robustness against the mobility can be guaranteed for both BPSK and DSK schemes, thanks to the property of ODDM modulation \cite{ODDM_tutorial}. However, the BER of BPSK increases rapidly with $K_{u}$ increasing, whose BER is more than $1\times10^{-2}$ under $P_{u}=4$ and $K_{u}=10$ even though the LMMSE detector can be deployed. The reason behind it is the growing MUI brought by the ever-decreasing guard interval when more UEs are involved. Fortunately, it can be mitigated significantly if DSK is applied, where the BER is less than 2\% of that in BPSK-based ODDM systems with $K_{u}=10$. It demonstrates the value of adopting DSK in uplink transmission again since both the PAPR and reliability can be ensured without loss of efficiency, which are two essential issues for UEs. Meanwhile, the maximum number of UEs that BPSK-based ODDM systems can support is $\lfloor\frac{N}{\log_{2}N}\rfloor=10$ if not sacrificing the achievable rate compared with DSK systems. However, this is not the limitation if DSK is employed, since the number of available sequences is $32$. It demonstrates the potential of uplink DSK frameworks for massive connectivity considering the ultra-reliable performance under $K_{u}=10$. At last, our proposed uplink DSK frameworks outperform BPSK-enabled DSK systems under arbitrary parameter selection in Fig. \ref{SimuFig_MU_Ku}, which finally verifies that DSK is qualified to address the uplink approach for emerging high-mobility applications in the 6G communication networks. \par 
\begin{figure}
	\centering{\includegraphics[width=0.83\linewidth]{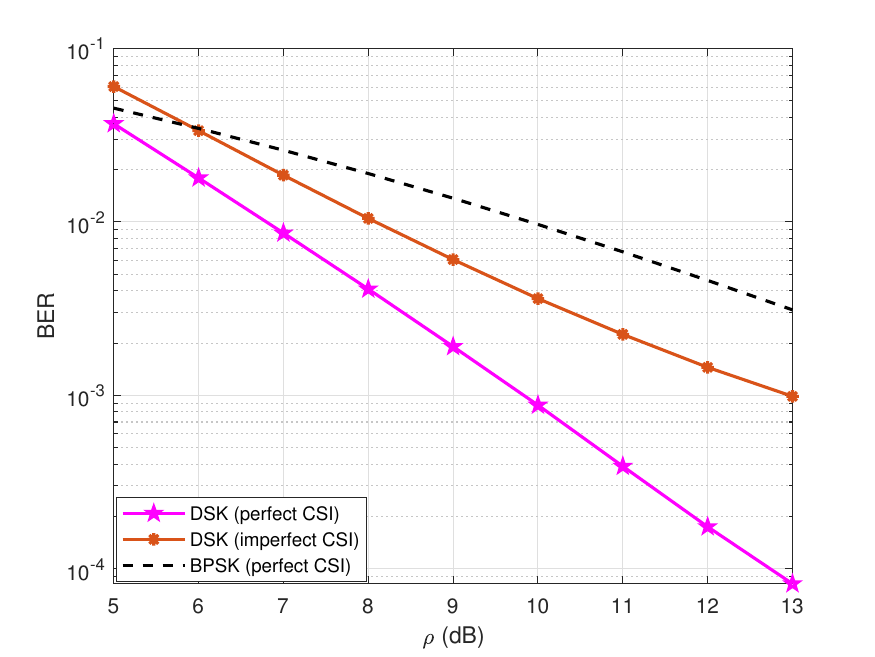}}
	\caption{BER against $\rho$ under imperfect CSI and $K_{u}=8$.}
	\label{SimuFig_MU_imperfect}	
\end{figure}
Finally, the performance evaluation with imperfect channel state information (CSI) is illustrated in Fig. \ref{SimuFig_MU_imperfect}, where we set $K_{u}=8$, $P_{u}=4$, and $v_{\max}=360$ km/h, respectively. Similar to the configurations in \cite{OTFS_SU_MPA,ODDM_imperfect_TCOM}, we adopt the estimation error as $(h_{p}^{u})^{\prime}=h_{p}^{u}+\Delta_{p}^{u}$, where we have $\Delta_{p}^{u}\sim\mathcal{CN}(0,\sigma_{e}^{2})$ with $\sigma_{e}^{2}=-20$ dB. It is obvious that when $\rho\geq6$ dB, DSK systems with imperfect CSI outperform BPSK schemes with perfect CSI. When $\rho=13$ dB, the proposed DSK system can achieve the BER of less than $10^{-3}$ even considering the channel estimation errors. In conclusion, the results in Fig. \ref{SimuFig_MU_imperfect} can demonstrate the applicability of the proposed DSK framework, while further optimization and evaluation considering the channel estimation can be expected in future research.\par 
\subsection{Discussions}
\label{subsec_discuss}
In this subsection, we briefly discussed the major limitation of this work, which motivates potential future research directions. At first, we aim at developing the basic principle and sequence optimization for DSK modulation, which can reduce the PAPR and MUI for uplink DD domain transmission. Therefore, we consider the on-grid channel throughout this paper for simplicity, i.e., the normalized time delay $l_{p}^{u}=\frac{\tau_{p}^{u}}{\tau_{r}}$ and Doppler shift $k_{p}^{u}=\frac{\nu_{p}^{u}}{\nu_{r}}$ are assumed to be integers. We acknowledge that general physical channels might involve off-grid components, which could significantly destroy the sparsity of ESDD channels \cite{ODDM_ESDD_ref_jun_TCOM}. It is apparent that the proposed DSK framework still works since the circular convolution along ODDM subcarriers can be satisfied by setting an appropriate ZP range \cite{Lowcomdet_ODDM_ZP}. However, the corresponding detection scheme requires adjustment to avoid the error propagation brought by the reduced sparsity in ESDD channels, e.g., by merging the subcarriers within an ODDM multicarrier symbol to reduce propagation paths in the factor graph like \cite{Lowcomdet_ODDM_ZP}. This will be embodied in our future research to further verify the superiority of the proposed DSK scheme in practical channels.\par 
On the other hand, we focus on the transmission frame and the corresponding detection design over uplink doubly-dispersive channels. For ease of the system analysis and illustration, the other procedures within the communications are assumed to be ideal. We acknowledge that even though Fig. \ref{SimuFig_MU_imperfect} has demonstrated the performance with channel estimation errors, some other practical elements might still affect the reliability, e.g., imperfect synchronization and other non-ideal effects brought by the hardware constraints. There is no doubt that the consideration of these elements can lead to a better system characterization and be more suitable for practical implementation, which will be possibly embodied in our future research.\par 
Finally, the major novelty of the proposed DSK framework lies in how the information bits are mapped into DD domain components, rather than how the DD domain components are transferred into the continuous-time waveform. Therefore, the proposed framework can be simply integrated into any kind of DD domain modulation system. Considering the lower OOBE and implementation complexity of ODDM than OTFS reported in \cite{ODDM_ESDD_ref_jun_TCOM,ref_ODDM,ODDM_tutorial}, we select ODDM to realize the DD domain modulation. Future research may consider a comprehensive comparison between ODDM and OTFS systems, where the recent progress of Zak-OTFS can be considered \cite{OTFS_2_BITS,OTFS_2_BITS_arxiv}. \par
\section{Conclusion}
\label{sec:conclusion}
In this paper, the ODDM modulation-based DSK framework is proposed to address the PAPR and reliability-complexity trade-off for doubly-dispersive channels. The framework of DSK transceivers is first established for point-to-point transmission. By generalizing the DSK modulation to the ZAC sequence-based one, the discussion is extended to uplink scenarios with the iterative SIC-MRC detector. The optimization of MU basis sequences is then provided by employing ZC sequences and analyzing the property of roots allocated to each UE. Simulation results demonstrate that our proposed framework enjoys ultra-low PAPR and better BER performance with the low-complexity detector, which indicates the potential of DSK to enhance the transmission performance under high-mobility scenarios. Future research directions will consider the corresponding diversity analysis and the integration of MIMO systems to promote the system throughput.\par 
\appendices
\section{Proof of Lemma \ref{lemma1_cross_correlation}}
\label{lemma1_proof}
At first, $\psi_{u,v}(k_{u},k_{v})$ can be computed as
\begin{equation}
	\begin{aligned}
		&\psi_{u,v}(k_{u},k_{v})=\sum_{k=0}^{N-1}z_{u}^{*}[(k-k_{u})_{N}]z_{v}[(k-k_{v})_{N}]\\
		&=\frac{1}{N}\sum_{k=0}^{N-1}e^{-j\pi\frac{m_{u}(k-k_{u})^{2}}{N}}e^{j\pi\frac{m_{v}(k-k_{v})^{2}}{N}}\\
		&=\frac{1}{N}\sum_{k=0}^{N-1}e^{j\pi\frac{(m_{v}-m_{u})k^{2}+2(k_{u}m_{u}-k_{v}m_{v})k+m_{v}k_{v}^{2}-m_{u}k_{u}^{2}}{N}}\\
		&=e^{j\pi\frac{m_{v}k_{v}^{2}-m_{u}k_{u}^{2}}{N}}\times\frac{1}{\sqrt{N}}\sum_{k=0}^{N-1}F[k]e^{j2\pi\frac{(k_{u}m_{u}-k_{v}m_{v})k}{N}},
	\end{aligned}
\label{lemma1_proof_first}
\end{equation}
where we have $F[k]=\frac{1}{\sqrt{N}}e^{j\pi(m_{v}-m_{u})\frac{k^{2}}{N}}$ for $k=0,1,\cdots,N-1$. Let $f[n]$ represent the normalized IDFT of $F[k]$, \eqref{lemma1_proof_first} can be simplified as
\begin{equation}
	|\psi_{u,v}(k_{u},k_{v})|^{2}=|f[(k_{u}m_{u}-k_{v}m_{v})_{N}]|^{2}.
	\label{eq_cross_IDFT}
\end{equation}
The main task can then be translated into seeking out $|f[n]|^{2}$. To solve this problem, let $R_{F}[m]=\sum_{k=0}^{N-1}F^{*}[(k-m)_{N}]F[k]$ represent the auto-correlation of $F[k]$, which has a close relation\footnote{The relation can be easily obtained by employing the property of DFT and IDFT.} to $|f[n]|^{2}$ as
\begin{equation}
	|f[n]|^{2}=\frac{1}{N}\sum_{m=0}^{N-1}R_{F}[m]e^{j2\pi\frac{mn}{N}}.
\end{equation}
Therefore, the auto-correlation is first simplified as
\begin{equation}
	\begin{aligned}
		R_{F}[m]&=\sum_{k=0}^{N-1}F^{*}[(k-m)_{N}]F[k]\\
		&=\frac{1}{N}\sum_{k=0}^{N-1}e^{j\pi(m_{v}-m_{u})\frac{k^{2}-(k-m)^{2}}{N}}\\
		&=\frac{e^{-j\pi(m_{v}-m_{u})\frac{m^{2}}{N}}}{N}\sum_{k=0}^{N-1}e^{j2\pi\frac{km(m_{v}-m_{u})}{N}}\\
		&=e^{-j\pi(m_{v}-m_{u})\frac{m^{2}}{N}}\delta[\left(m(m_{v}-m_{u})\right)_{N}].
	\end{aligned}
\label{simplify_autocor}
\end{equation}
When $\left((m_{v}-m_{u})m\right)_{N}\ne0$, it can be directly deduced as $R_{F}[m]=0$. When $m=0$ or $m_{u}=m_{v}$, \eqref{simplify_autocor} can be computed as $1$. For other scenarios, we have $\left((m_{v}-m_{u})m\right)_{N}=0$ with $m\ne0$ and $m_{v}\ne m_{u}$, i.e., $(m_{v}-m_{u})=\frac{\gamma N}{m}$ with $\gamma\in\mathbb{Z}\backslash\{0\}$ holds. On the other hand, $\left((m_{v}-m_{u})m\right)_{N}=0$ is equivalent to $\text{gcd}((m_{v}-m_{u})m,N)=N=2^{n}$. Since $m_{u}\in[1,N-1]$ holds for $\forall~u$, we can deduce $\text{gcd}(m_{v}-m_{u},N)\leq2^{n-1}$, which indicates $\text{gcd}(m,N)\geq2$ must be satisfied. Considering all factors of $N$ can be derived as $2^{n^{\prime}}$ with $n^{\prime}\in\mathbb{Z}$ and $0\leq n^{\prime}\leq n$, $m$ is certainly even. Therefore, it can be derived as
\begin{equation}
	\begin{aligned}
		e^{-j\pi(m_{v}-m_{u})\frac{m^{2}}{N}}=e^{-j\pi\gamma m}=1,
	\end{aligned}
\end{equation}
which leads to $R_{F}[m]=\delta[(m(m_{v}-m_{u}))_{N}]$ by combining the analysis in \eqref{simplify_autocor}. For ease of illustration, the notations $c=\text{gcd}(m_{v}-m_{u},N)$ and $\lambda=\frac{N}{c}$ are adopted. Then $R_{F}[m]$ can be further simplified as
\begin{equation}
	R_{F}[m]=\begin{cases}
		1,&(m)_{\lambda}=0\\
		0,&\text{elsewhere}
	\end{cases}.
\end{equation}
Consequently, $|f[n]|^{2}$ can be derived as
\begin{equation}
	\begin{aligned}
		|f[n]|^{2}&=\frac{1}{N}\sum_{m=0}^{N-1}R_{F}[m]e^{j2\pi\frac{mn}{N}}\\
		&=\frac{1}{N}\sum_{c_{1}=0}^{c-1}e^{j2\pi\frac{c_{1}\lambda n}{N}}=\frac{1}{N}\sum_{c_{1}=0}^{c-1}e^{j2\pi\frac{c_{1}n}{c}}\\
		&=\begin{cases}
			\frac{c}{N},&(n)_{c}=0\\
			0,&\text{elsewhere}
		\end{cases}.
	\end{aligned}
\end{equation}
At last, combining the derivation in \eqref{eq_cross_IDFT}, $|\psi_{u,v}(k_{u},k_{v})|^{2}$ can be provided by
\begin{equation}
	\begin{aligned}
		&|\psi_{u,v}(k_{u},k_{v})|^{2}\\
		&=\begin{cases}
			\frac{c}{N},&\left((k_{u}m_{u}-k_{v}m_{v})_{N}\right)_{c}=0\\
			0,&\text{elsewhere}	
		\end{cases}.
	\end{aligned}
\end{equation}
which completes the proof of \textbf{Lemma \ref{lemma1_cross_correlation}}.
\section{Proof of Theorem \ref{theorem_MUsequences}}
\label{theorem_MUsequences_proof}
Since $\text{gcd}(m_{u},N)=1$ is required considering the property of auto-correlation, $m_{u}$ must be odd for $\forall~u$. In this appendix, we first provide the analysis of the solution $m_{u}=2u+m_{0}$. Then it will be proved that any other solution cannot outperform it. Since the scenarios when $K_{u}\leq2$ are trivial, we focus on the derivation with $2<K_{u}\leq2^{n-1}=\frac{N}{2}$, where we adopt the notation as $m=\lceil\log_{2}K_{u}\rceil-1$ with $1\leq m<n-1$ and $Q=K_{u}-2^{m}$. \par 
Considering the property of arithmetic sequences, $|m_{u}-m_{v}|$ for $u\ne v$ can be any even number within $[2,2K_{u}-2]$, which indicates the maximum value of $\text{gcd}(m_{u}-m_{v},N)$ is $2^{m+1}$ since $2^{m+1}\leq2K_{u}-2<2^{m+2}$. At the meantime, this means $\text{gcd}(m_{u}-m_{v},N)$ reaches $2^{m+1}$ when $u-v=2^{m}$. Therefore, there are $Q$ pairs of UE corresponding to the maximum value of $\text{gcd}(m_{u}-m_{v},N)$ as $(0,2^{m}), (1,2^{m}+1), \cdots, (Q-1,2^{m}+Q-1)$. In conclusion, for the proposed selection, the maximum value of $\text{gcd}(m_{u}-m_{v},N)$ is $2^{m+1}$ with $Q$ accessible pairs.\par 
Then we prove the optimality of the proposed solution. For each selection, let $m_{u}=\sum_{t=0}^{n-1}\text{b}_{u}^{t}2^{t}$ represent the binary representation of $m_{u}$ with $\text{b}_{u}^{t}\in\{0,1\}$. $\text{b}_{u}^{0}=1$ is required to guarantee $m_{u}$ is relatively prime to $N=2^{n}$. Since we have $2^{m}<K_{u}$, there is at least one pair of $(u,v)$ with $u\ne v$ and $\text{b}_{u}^{t}=\text{b}_{v}^{t}$ for $\forall~0\leq t\leq m$ according to the \textit{pigeonhole principle}. Then for this pair of $(u,v)$, it can be derived that $(m_{u}-m_{v})_{2^{m+1}}=0$, which indicates $\mathop{\max}_{u,v}\text{gcd}(m_{u}-m_{v},N)\geq2^{m+1}$ holds for any possible choices of $\{m_{u}|u=0,1,\cdots,K_{u}-1\}$. Since the proposed choice holds $\mathop{\max}_{u,v}\text{gcd}(m_{u}-m_{v},N)=2^{m+1}$, it is the optimal method when it comes to the maximum value of $\text{gcd}(m_{u}-m_{v},N)$.\par 
Assuming that there is a strategy $\{m_{u}|u=0,\cdots,K_{u}-1\}$ satisfying only $Q_{1}<Q$ UE pairs\footnote{The scenario for $Q=1$ is trivial, which is not included in the discussion.} reach $\text{gcd}(m_{u}-m_{v},N)=2^{m+1}$. It indicates that there exists a subset with $K_{u}-Q_{1}>2^{m}$ elements as $\mathcal{U}\subsetneqq \{0,1,\cdots,K_{u}-1\}$, where $\left\{\text{b}_{u}^{t}=\text{b}_{v}^{t}, \forall~0\leq t\leq m\right\}$ does not hold for $\forall~u\ne v$ and $u,v\in\mathcal{U}$. However, according to the \textit{pigeonhole principle}, there is certainly a pair of $u,v\in\mathcal{U}$ with $\text{b}_{u}^{t}=\text{b}_{v}^{t}$ for $\forall~0\leq t\leq m$ since $|\mathcal{U}|>2^{m}$. It is contradictory from the basic assumption of $\mathcal{U}$, which reveals that there does not exist a strategy with only $Q_{1}<Q$ pairs of $(u,v)$ that reaches $\text{gcd}(m_{u}-m_{v},N)=2^{m+1}$. On the other hand, there are only $Q$ reachable pairs for the proposed strategy $m_{u}=2u+m_{0}$, which is already the metric bound.\par 
Based on the aforementioned analysis, the optimality of the proposed selection of roots as $m_{u}=2u+m_{0}$ is successfully demonstrated, which completes the proof of \textbf{Theorem \ref{theorem_MUsequences}}.
\bibliographystyle{IEEEtran}
\bibliography{ref-sum}

\end{document}